\documentclass[english]{achemso}
\usepackage{geometry}
\usepackage{booktabs}
\usepackage{textcomp}
\usepackage{mathtools}
\usepackage{amsmath}
\usepackage{graphicx}
\usepackage{multirow}
\usepackage{makecell}
\usepackage{xcolor}

\usepackage{ifxetex}
\ifxetex
  \usepackage{fontspec}
\else
  \usepackage[T1]{fontenc}
  \usepackage[latin9]{inputenc}
  \usepackage{lmodern}
\fi

\makeatletter

\title{Molecular resonance identification in complex absorbing potentials
via integrated quantum computing and high-throughput computing}
\author{Jingcheng Dai,$^{\dagger,a}$ Atharva Vidwans,$^{\dagger,a,b}$ Eric
H.~Wan,$^{c,d}$ Alexander X.~Miller, and Micheline B.~Soley$^{a,b,e}$}
\affiliation{$^{a}$Department of Chemistry, University of Wisconsin-Madison,
1101 University Avenue, Madison, WI 53706, USA\\
$^{b}$Department of Physics, University of Wisconsin-Madison, 1150
University Avenue, Madison, WI 53706, USA\\
$^{c}$Department of Mathematics, University of Wisconsin-Madison,
480 Lincoln Drive, Madison, WI 53706, USA\\
$^{d}$Department of Computer Sciences, University of Wisconsin-Madison,
1210 West Dayton Street, Madison, WI 53706, USA\\
$^{e}$Data Science Institute, University of Wisconsin-Madison, 447
Lorch Street, Madison, WI 53706, USA\\
$^{\dagger}$These authors contributed equally to this work.\\
$^{*}$Corresponding Author}
\email{msoley@wisc.edu}
\providecommand{\tabularnewline}{\\}

\setkeys{acs}{articletitle=true}

\makeatother

\usepackage{babel}
\begin{document}
\newpage{}
\begin{abstract}
Recent advancements in quantum algorithms have reached a state where
we can consider how to capitalize on quantum and classical computational
resources to accelerate molecular resonance state identification.
Here we identify molecular resonances with a method that combines
quantum computing with classical high-throughput computing (HTC).
This algorithm, which we term qDRIVE (the quantum deflation resonance
identification variational eigensolver) exploits the complex absorbing
potential formalism to distill the problem of molecular resonance
identification into a network of hybrid quantum-classical variational
quantum eigensolver tasks, and harnesses HTC resources to execute
these interconnected but independent tasks both asynchronously and
in parallel, a strategy that minimizes wall time to completion. We
show qDRIVE successfully identifies resonance energies and wavefunctions
in simulated quantum processors with current and planned specifications,
which bodes well for qDRIVE's ultimate application in disciplines
ranging from photocatalysis to quantum control and places a spotlight
on the potential offered by integrated heterogenous quantum computing/HTC
approaches in computational chemistry.\medskip{}

\qquad{}\includegraphics[width=0.75\textwidth]{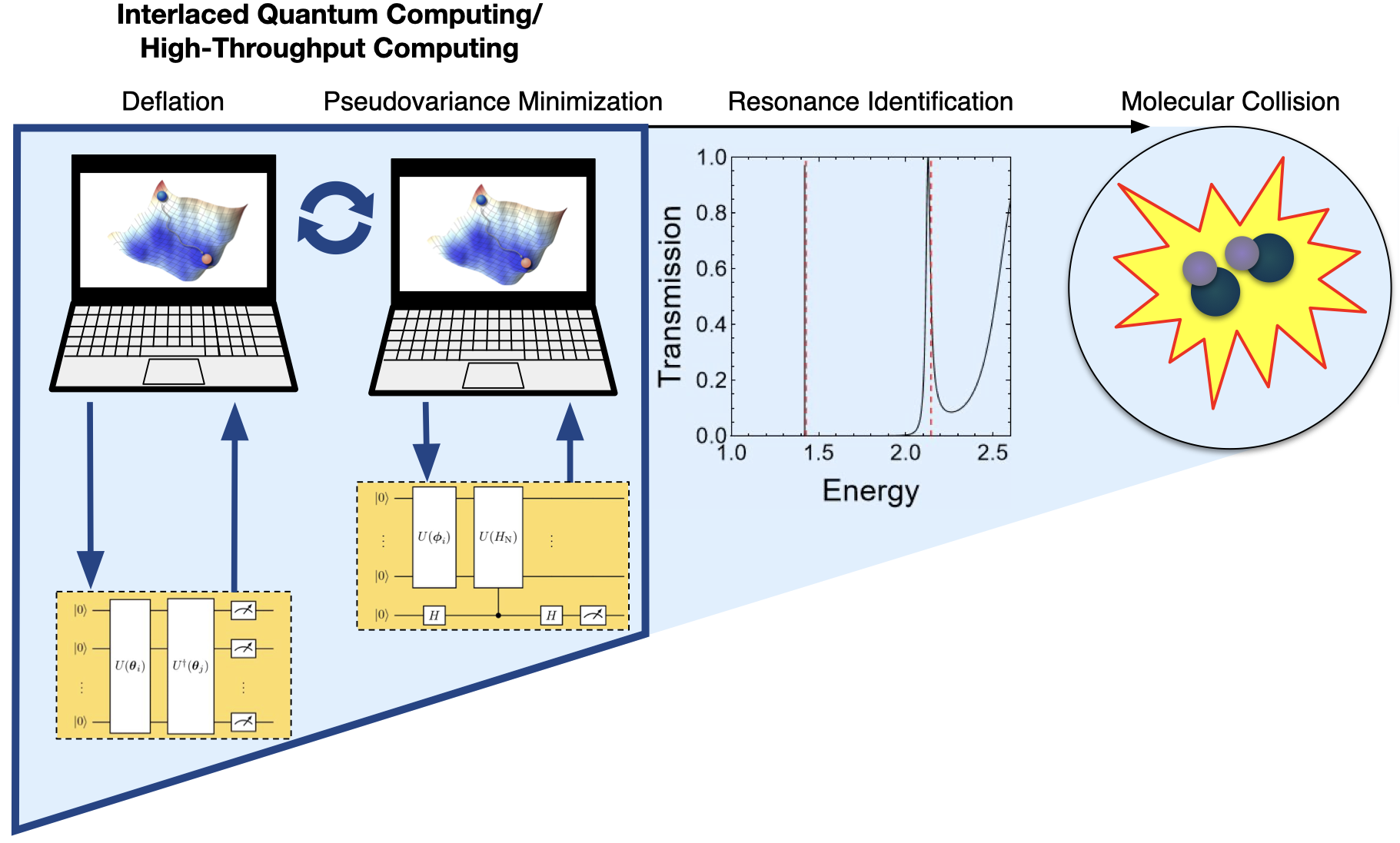}
\end{abstract}

\section{Introduction}

To date, much of the focus on quantum algorithms in chemistry has
been placed on identification of the ground state and excited states
of Hermitian Hamiltonians.\cite{abrams1999quantum,farhi2000quantum,peruzzo2014variational,albash2018adiabatic,motta2020determining,kyaw2023boosting,hejazi2024adiabatic,santagati2018witnessing,colless2018computation,higgott2019variational,nakanishi2019subspace,ollitrault2020quantum,kuroiwa2021penalty,kim2023two,wang2023electronic,cianci2024subspace,grimsley2025challenging}
 However, an emerging interest lies in open quantum systems, both
to simulate realistic chemical systems in contact with surrounding
environments\cite{delgado2025quantum} and to characterize physical
systems --- including molecular qubits --- to use as the basis of novel
quantum computing architectures.\cite{wasielewski2020exploiting}
The newfound interest in open quantum systems has led to the development
of quantum algorithms to identify eigenstates of associated non-Hermitian Hamiltonians,\cite{wang2010measurement,daskin2014universal,parker2020quantum,shao2020quantum,teplukhin2020solving,shao2022computing,zhao2023universal,singh2024quantum,xie2024variational,hancock2025quantum}
including left- and right-eigenvectors associated with parity-time
reversal ($\mathcal{PT}$) symmetry breaking at exceptional points\cite{xie2024variational,hancock2025quantum}
and resonances (purely outgoing eigenstates) associated with decay
processes.\cite{daskin2014universal,teplukhin2020solving,singh2024quantum,singh2025quantum}

Quantum algorithms for resonance identification are of particular
relevance in today's computational chemistry since resonances are
ubiquitous for the description of molecular breakup processes in chemistry
ranging from ultracold collision complex decay\cite{park2023feshbach}
to plasmonic photocatalysis.\cite{Jagau2017ExtendingQC} Recent predictions
also suggest molecular resonances could play a key role in the ongoing
second quantum revolution with impacts to both quantum information
processing\cite{cornish2024quantum} and quantum control.\cite{son2022control}
Many classical methods to identify resonances exist, such as approaches
based on complex absorbing potentials (CAPs),\cite{kosloff1984dynamical,jolicard1985optical,kosloff1986absorbing,neuhasuer1989time,Muga2004Complex}
complex scaling,\cite{aguilar1971class,balslev1971spectral} Feshbach
projection,\cite{livshits1957application,feshbach1958unified,feshbach1962unified}
the stabilization method,\cite{taylor1966qualitative,eliezer1967resonant,taylor1970models}
analytic continuation in the coupling constant,\cite{kukulin1977description,kukulin1979method}
and R-matrix theory\cite{descouvemont2024resonances}. However, contemporary
methods for molecular resonance identification often struggle with
exacting parameter dependence and/or sizeable basis sets.\cite{Jagau2017ExtendingQC}
This begets the question: how can we harness available computational
resources --- both quantum and classical --- to facilitate and accelerate
molecular resonance identification?

Current quantum algorithms that address this question tend to fall
into two categories.\footnote{Note an outlier, the Quantum Annealer Eigensolver,\cite{teplukhin2020solving}
which is specifically designed for use on quantum annealing (e.g.,
D-Wave) systems.\cite{yarkoni2022quantum}} The first category comprises algorithms based on quantum phase estimation.
This category can be considered to include algorithms such as measurement-based
quantum phase estimation,\cite{wang2010measurement} iterative phase
estimation,\cite{daskin2014universal} generalizations of quantum
phase estimation,\cite{parker2020quantum,shao2020quantum,shao2022computing}
the iterative Harrow-Hassidim-Lloyd approach,\cite{zhang2025iterative}
and the direct measurement method.\cite{bian2019quantum,zhang2025quantum,singh2024quantum}
Since quantum phase estimation tends to call for a high circuit depth
(gate count), algorithms in this category are often best suited for
fault-tolerant quantum computers still under development. The second
category comprises algorithms that adapt the variational quantum eigensolver
(VQE)\cite{peruzzo2014variational,cao2019quantum,mcardle2020quantum,cerezo2021variational,Tilly2022Variational}
to non-Hermitian systems. This category can be considered to include
algorithms such as the Variational Quantum Universal Eigensolver\cite{zhao2023universal}
and Xie-Xue-Zhang method\cite{xie2024variational,hancock2025quantum,singh2025quantum}.
Since these algorithms rely on a VQE-like approach, algorithms in
this category typically require lower circuit depths than algorithms
based on quantum phase estimation, and thus are more naturally suited
to existing noisy intermediate scale quantum (NISQ) computers.
The adaptation of VQE to non-Hermitian systems entails an added computational
cost related to preparation of the Ansatz (trial or guess state);
for example, the Variational Quantum Universal Eigensolver requires
preliminary precise training\cite{zhao2023universal} and the Xie-Xue-Zhang
method requires both parametrization of and scans over the eigenenergy.\cite{xie2024variational}

In this study, we facilitate molecular resonance identification by
combining the complex absorbing potential (CAP) formalism with an
interlaced quantum computing/parallel asynchronous high-throughput
computing (HTC) approach that we term qDRIVE (the quantum deflation resonance
identification variational eigensolver). 

We recognize that the construction of CAPs enables us to prepare a
physically motivated Ansatz without precise training or energy parametrization
as follows: In the CAP formalism,\cite{leforestier1983optical,kosloff1984dynamical,jolicard1985optical,leforestier1985role,kosloff1986absorbing,neuhasuer1989time,Muga2004Complex,Jagau2017ExtendingQC}
resonances are identified by appending to the original potential of
interest $V_{0}$ a negative imaginary potential at the outer reaches
of the simulation window $\text{i}V_{\text{CAP}}$ to impose purely
outgoing boundary conditions. The CAP formalism thus entails two Hamiltonians:
a Hermitian Hamiltonian $H_{\text{H}}$ comprised of the system's
physical kinetic and potential energy and an artificial non-Hermitian
Hamiltonian that is only made non-Hermitian by the imposition of boundary
conditions with a CAP $H_{\text{N}}$. Since the two Hamiltonians
differ only at the extremities of the simulation window, their eigenstates
are, in many systems, likely to be similar in the internal region.
Therefore, an effective initial guess for an eigenstate of the non-Hermitian
Hamiltonian is often the corresponding eigenstate of the Hermitian
Hamiltonian. qDRIVE uses this strategy to prepare the initial Ansatz
for the non-Hermitian Hamiltonian resonance as the corresponding eigenstate
of the Hermitian Hamiltonian. This identification of the required
Hermitian Hamiltonian eigenstates is a straightforward task on quantum
computers with VQE\cite{peruzzo2014variational,cao2019quantum,mcardle2020quantum,cerezo2021variational,Tilly2022Variational}
and its excited-state analog, variational quantum deflation (VQD).\cite{higgott2019variational}

\textcolor{black}{Importantly, CAP-based qDRIVE provides the means to sidestep many of the complexities involved in non-Hermitian quantum mechanics. Specifically, in Hermitian quantum mechanics, wavefunctions are decomposed in terms of the complete set of eigenfunctions of a Hermitian operator, where orthogonality is determined according to the standard scalar product $\left\langle \psi\middle|\phi\right\rangle =\int\text{d}x\,\psi^{\star}\left(x\right)\phi\left(x\right)$. In non-Hermitian quantum mechanics, decomposition of wavefunctions instead requires the c-product given by $\left(\psi\middle|\phi\right)=\int\text{d}x\,\psi\left(x\right)\phi\left(x\right)$.\cite{herzenberg1963resonant,moiseyev1978resonance} Here a significant complication arises -- eigenvalues of a non-Hermitian operator can coalesce, as often occurs in parity-time-reversal-symmetric ($\mathcal{PT}$-symmetric) systems\cite{konotop2016nonlinear,miri2019exceptional,ozdemir2019parity} and in complex scaling at the complex rotation angle where continuum and resonance solutions meet.\cite{moiseyev1980association} In this case, eigenvalues become defective and exceptional points appear; and  eigenfunctions themselves coalesce such that the spectrum becomes incomplete or defective. Simulation methods must then contend with self-orthogonality, in which the c-product becomes zero, and an accompanying lack of biorthogonal completeness. Such a situation typically calls for special adaptations such as the inclusion of functions beyond true eigenfunctions in order to ensure closure relations hold.\cite{Moiseyev2011NonHermitian}}

\textcolor{black}{CAP-based resonance identification in qDRIVE allays these concerns at two levels: Firstly, at the algorithmic level, by construction, qDRIVE never calls for calculation of the c-product, only the standard scalar product, for which these problems do not occur. 
Secondly, at the NISQ computing level, inherent quantum processor error is expected to break apart degeneracies. It is well known that an infinitesimal perturbation is sufficient to destroy an exceptional point: for example, round-off error is sufficient to avoid a zero c-product in finite-element approaches.\cite{moiseyev1980association,Moiseyev2011NonHermitian} By definition, NISQ computers similarly involve a significant (typically larger) degree of error, both in terms of thermal and gate noise and in terms of limitations on the gate count that lead to finite Ansatz expressibility. Just as round-off error pushes systems away from the exceptional point on classical processors, NISQ systems' error is similarly expected to drive systems away from exact exceptional points to allow for the restoration of biorthogonal completeness.}

Today, there is active discussion on how to employ high-performance
computing (HPC) resources to accelerate quantum information processing
tasks.\cite{barral2025review,elsharkawy2025integration,rallis2025interfacing,mansfield2025first}
Traditionally, such HPC approaches accelerate computation by running
blocks of jobs in parallel, often with tight communication between
jobs. However, where HPC resources are limited, such an approach can
require large wait times until a block of a suitable size becomes
available. In contrast, HTC follows a vulture's approach that scavenges
individual available computational cores as they become available
to run jobs --- still in parallel, but now asynchronously as resources
allow.\cite{litzkow1987remote,litzkow1987condor,livny1997mechanisms,basney1999deploying,thain2005distributed,morgan2009high,fajardo2015much}
This HTC approach maximizes throughput of computational tasks performed
by the supercomputer over time, an important tack where supercomputing
resources are scarce. qDRIVE consists of a network of interconnected
but independently executable jobs. This network can be readily expressed
as a directed acyclic graph, which can be implemented on HTC resources
with HTCondor DAGMan (the HTCondor directed acyclic graph manager).\cite{litzkow1987remote,litzkow1987condor,livny1997mechanisms,basney1999deploying,thain2005distributed,morgan2009high,fajardo2015much}

\textcolor{black}{Typically, HTCondor DAGMan uses fall into three main categories based on the instructed order of task operations: (1) sequential DAGs, in which task A is completed prior to task B, task B prior to task C, and so on; (2) split-and-recombine DAGs, in which a task begets many tasks that feed into a single shared task; and (3) collection DAGs, in which there is no ordering of tasks to be completed.\cite{osgconnectgithub} qDRIVE uses a DAG that falls outside these categories. This DAG can be described rigorously as an arborescence in which each parent node leads to a child node of the same connectivity and a child node with no children. Such an interlaced DAG hybridizes the cascading nature of a sequential DAG with the spawning ability of a split-and-recombine DAG. In addition, the DAG itself is executed in separate batch runs, such that the DAG also shares the embarrassingly parallel characteristic nature of a collection DAG. This innovative hybrid DAG approach accelerates completion of all of qDRIVE's interdependent Hermitian and non-Hermitian Hamiltonian stages by closely mirroring qDRIVE's algorithmic structure and the execution of tasks with the DAGMan metascheduler.} 

\textcolor{black}{To the knowledge of the authors, the framework introduced in qDRIVE is the only framework to harness HTC resources in quantum computing for chemical applications to date in the literature. The only known use of comparable high-throughput computing systems in quantum computing is the use of HTCondor to simulate quantum computers classically,\cite{isakov2021simulations} which constitutes a distinct goal from the acceleration of hybrid quantum-classical algorithms that may employ real quantum computers as put forward here. The HTC framework introduced in qDRIVE is therefore expected to be of unique benefit to researchers seeking to accelerate completion of hybrid quantum-classical algorithm jobs, as well as to those who have access to publicly available HTC resources such as the Open Science Grid but who may have only limited access to more widely known HPC resources.\cite{pordes2007open,altunay2011science}}

In proof-of-concept experiments using a variety of quantum simulators,
we show qDRIVE identifies the bound and resonance states of a long-established
benchmark model of molecular predissociation\cite{moiseyev1978resonance,bian2019quantum}
(i) in the zero-noise limit and (ii) on NISQ systems in conjunction
with the practical error mitigation techniques\cite{endo2018practical,cai2023quantum}
of readout matrix inversion\cite{dewes2012characterization,maciejewski2020mitigation,geller2020rigorous}
and hybrid linear-exponential zero-noise extrapolation (ZNE)\cite{temme2017error,endo2018practical,kandala2019error,giurgica2020digital,miller2025universal}.
For the same system, we also show qDRIVE fares well (iii) in far-term
environments, modeled by the anticipated gate errors and qubit longevities
of quantum processors on the horizon. These successes show that qDRIVE
can be used to identify molecular resonance states capitalizing on
a joint quantum computing and HTC heterogenous computing approach.

\section{Methods}

\subsection{qDRIVE Algorithmic Framework}

As the basis of a quantum algorithm for molecular resonance identification
in the complex absorbing potential (CAP) formalism,\cite{leforestier1983optical,kosloff1984dynamical,jolicard1985optical,leforestier1985role,kosloff1986absorbing,neuhasuer1989time,Muga2004Complex,Jagau2017ExtendingQC}
we consider the class of Siegert pseudostates defined as the purely
outgoing eigenstates of a non-Hermitian Hamiltonian $H_{\text{N}}$
given by the sum of (i) the Hermitian Hamiltonian $H_{\text{H}}$
corresponding to the purely real potential $V_{0}$ of interest and
(ii) a complex absorbing potential (CAP) $V_{\text{CAP}}$
\begin{equation}
H_{\text{N}}=H_{\text{H}}+\text{i}V_{\text{CAP}}\label{eq:HsubN}
\end{equation}
where $V_{\text{CAP}}$ is purely a real potential that is zero in
the internal region of physical interest and negative near the simulation
boundary in order to impose purely outgoing boundary conditions. Eigenstates
of the resulting non-Hermitian Hamiltonian are complex in energy $E_{\text{}}=E_{\text{r}}-\text{i}E_{\text{i}}$
with an imaginary part that corresponds to the resonance decay rate
$E_{\text{i}}=\Gamma/2$ and lifetime $\tau=\Gamma^{-1}$. Note practical
identification of physical Siegert states $H_{\text{N}}\psi=E\psi$
further requires omission of spurious resonance states that correspond
to artifacts native to the CAP method (namely, nonresonant, diverging,
and indifferent states, see ref.~\citenum{Riss1993Calculation}).

To characterize resonances in this CAP formalism, we consider the
purely nonnegative metric \textcolor{black}{for a Hamiltonian $\text{H}_\text{O}$ of
\begin{equation}
\sigma_{\text{pseudo}}^{2}=\left\langle H_\text{O}^{\dagger}H_\text{O}\right\rangle -\left\langle H_\text{O}^{\dagger}\right\rangle \left\langle H_\text{O}\right\rangle ,\label{eq:Pseudovariance}
\end{equation}
where the subscript $\text{O}=\{\text{H},\text{N}\}$ denotes the Hermiticity or non-Hermiticity of the Hamiltonian, respectively, and where we refer to the metric as the pseudovariance to distinguish the
metric from the standard, simplified variance formula for Hermitian Hamiltonians $\text{O}=\text{H}$ used in standard Variance
VQE\cite{zhang2022variational} and related variance-based quantum algorithms for excited state determination such as the contracted quantum eigensolver (CQE),\cite{wang2023electronic,warren2024exact} VQE-X,\cite{zhang2021adaptive} CoVaR,\cite{boyd2022training} and excited-state variance VQE,\cite{hobday2022variance,liu2023probing} namely,
\begin{align}
\sigma^{2} & =\left\langle H_{\text{O}}^{2}\right\rangle -\left\langle H_{\text{O}}\right\rangle ^{2},
\end{align}
for $\text{O}=\text{H}$, which yields a nonordered field if applied to generic non-Hermitian
matrices $\text{O}=\text{N}$. This pseudovariance metric assumes the same form as the squared-residual cost function used in
variance VQE to identify right eigenvectors of non-Hermitian Hamiltonians\cite{xie2024variational} with the key exception that the pseudovariance analytically eliminates dependence on an energy parameter. Note that, since the pseudovariance}
obviates the need for an energy parameter, use of the pseudovariance circumvents
the parameter scan required by prior non-Hermitian Variance VQE approaches.\cite{xie2024variational,hancock2025quantum,singh2025quantum}
Furthermore, since the CAP-based Hamiltonian $H_{\text{N}}$ Eq.~(\ref{eq:HsubN})
is given by the sum of a Hermitian Hamiltonian $H_{\text{H}}$ and
a purely imaginary term $\text{i}V_{\text{CAP}}$, its expectation
value $\left\langle H_{\text{N}}\right\rangle $ follows directly
from the expectation values of two Hermitian operators, namely, $\left\langle H_{\text{N}}\right\rangle =\left\langle H_{\text{H}}\right\rangle +\text{i}\left\langle V_{\text{CAP}}\right\rangle $.
This decomposition allows the expectation value of the energy of a
resonance to be computed using only the scalar product\cite{Riss1993Calculation}
$\left\langle \psi\middle|\phi\right\rangle =\int\text{d}x\,\psi^{\star}\left(x\right)\phi\left(x\right)$,
without recourse to the standard c-product used for generic non-Hermitian
systems $\left(\psi\middle|\phi\right)=\int\text{d}x\,\psi\left(x\right)\phi\left(x\right)$.\cite{herzenberg1963resonant,moiseyev1978resonance}
Similar arguments hold for the remaining terms of the pseudovariance
$\left\langle H_{\text{N}}^{\dagger}\right\rangle $ and $\left\langle H_{\text{N}}^{\dagger}H_{\text{N}}\right\rangle $
such that (i) the pseudovariance may be computed using only the standard
scalar product and thereby its associated quantum circuits\cite{peruzzo2014variational,Tilly2022Variational}
without recourse to twin pairs of left- and right-eigenvector Ansatz
parameters\cite{xie2024variational,hancock2025quantum} and (ii) the
pseudovariance can be shown to be purely nonnegative and to feature
global minima of zero only for eigenstates, \emph{i.e.},
\begin{equation}
\sigma_{\text{pseudo}}^{2}\ge0
\end{equation}
where $\sigma_{\text{pseudo}}^{2}=0$ for states $\psi$ that satisfy
$H_{\text{N}}\psi=E\psi$.

\begin{figure}
\begin{centering}
\includegraphics[width=1\textwidth]{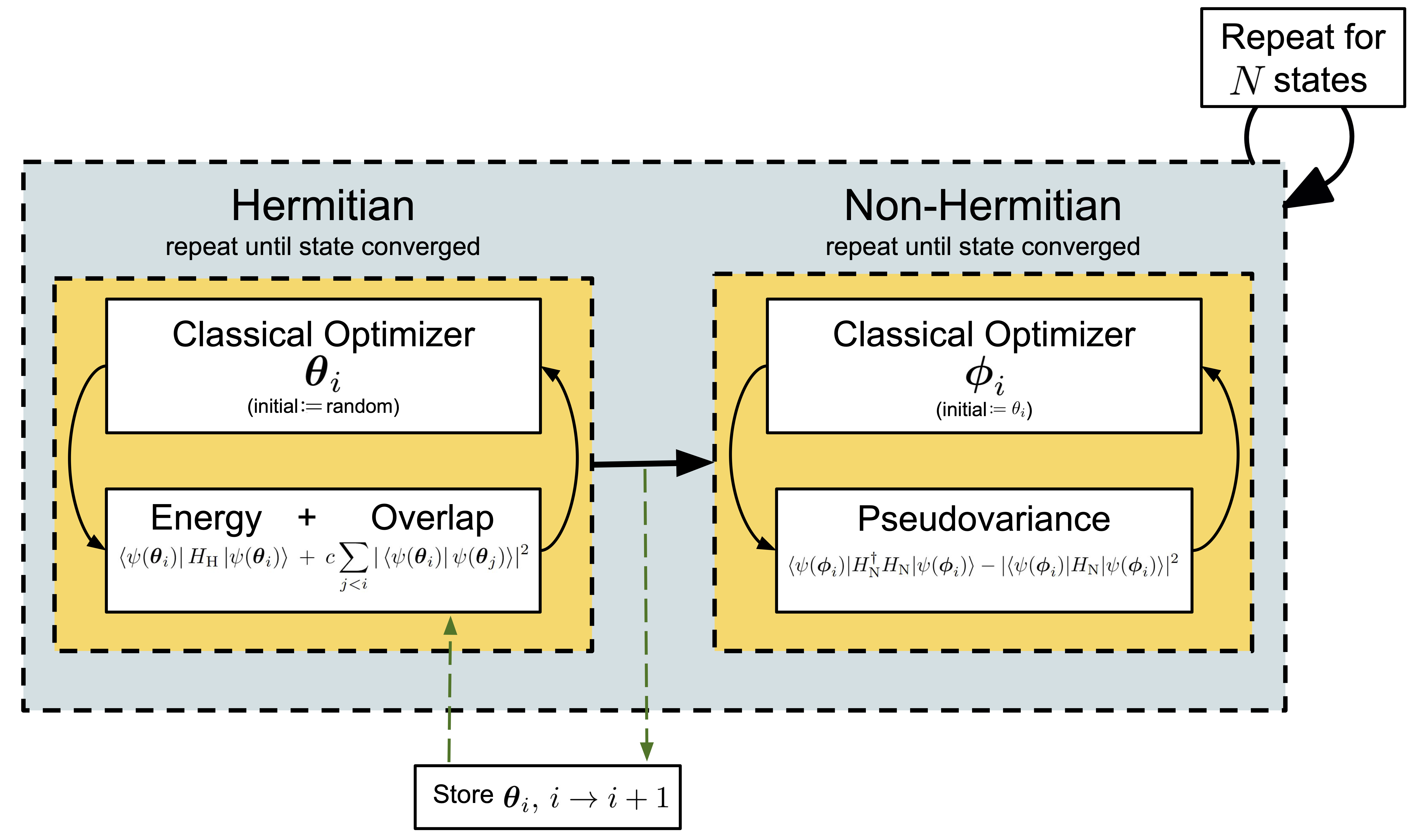}
\par\end{centering}
\caption{Schematic of the qDRIVE algorithm.\label{fig:Schematic}}
\end{figure}

These characteristics of the pseudovariance, combined with the aforementioned
close relationship between $H_{\text{N}}$ and $H_{\text{H}}$ in
the CAP formalism, enable the identification of $N$ eigenstates of
$H_{\text{N}}$ according to the scheme depicted in Fig.~\ref{fig:Schematic}
and outlined below:
\begin{enumerate}
\item Consider a parametrized Ansatz wavefunction $\psi\left(\mathbf{\theta}\right)$
subject to the CAP-based Hamiltonian $H_{\text{N}}=H_{\text{H}}+\text{i}V_{\text{CAP}}$
and initialize $i=1$ and a set of random parameters $\left\{ \mathbf{\theta}_{i}\right\} $.
\item For the Hermitian Hamiltonian $H_{\text{H}}$, determine the Ansatz
parameters $\mathbf{\theta}_{i}$ that correspond to the $i^{\text{th}}$
eigenstate via optimization of the objective function
\begin{equation}
\left\langle \psi\left(\mathbf{\theta}_{i}\right)\left|H_{\text{H}}\right|\psi\left(\mathbf{\theta}_{i}\right)\right\rangle +c\sum_{j<i}\left|\left\langle \psi\left(\mathbf{\theta}_{i}\right)\middle|\psi\left(\mathbf{\theta}_{j}\right)\right\rangle \right|^{2}\label{eq:VQDObjective}
\end{equation}
for penalty parameter $c$.
\item Store the optimized Ansatz parameters $\mathbf{\theta}_{i}$, initialize
$\mathbf{\phi}_{i}\coloneqq\mathbf{\theta}_{i}$, and increment $i\rightarrow i+1$.
In parallel:
\begin{enumerate}
\item Return to step 2\\
\\
AND\\
\item For the non-Hermitian Hamiltonian $H_{\text{N}}$, determine the Ansatz
parameters $\mathbf{\phi}_{i}$ that correspond to the $i^{\text{th}}$
eigenstate via optimization of the pseudovariance objective function
\begin{equation}
\left\langle \psi\left(\mathbf{\phi}_{i}\right)\left|H_{\text{N}}^{\dagger}H_{\text{N}}\right|\psi\left(\mathbf{\phi}_{i}\right)\right\rangle -\left|\left\langle \psi\left(\mathbf{\phi}_{i}\right)\left|H_{\text{N}}\right|\psi\left(\mathbf{\phi}_{i}\right)\right\rangle \right|^{2}.\label{eq:PseudovarianceObjective}
\end{equation}
\end{enumerate}
\item Proceed until $i=N+1$.
\end{enumerate}

\subsection{Hybrid Quantum-Classical Implementation}

\begin{figure*}
\begin{centering}
\includegraphics[width=1\textwidth]{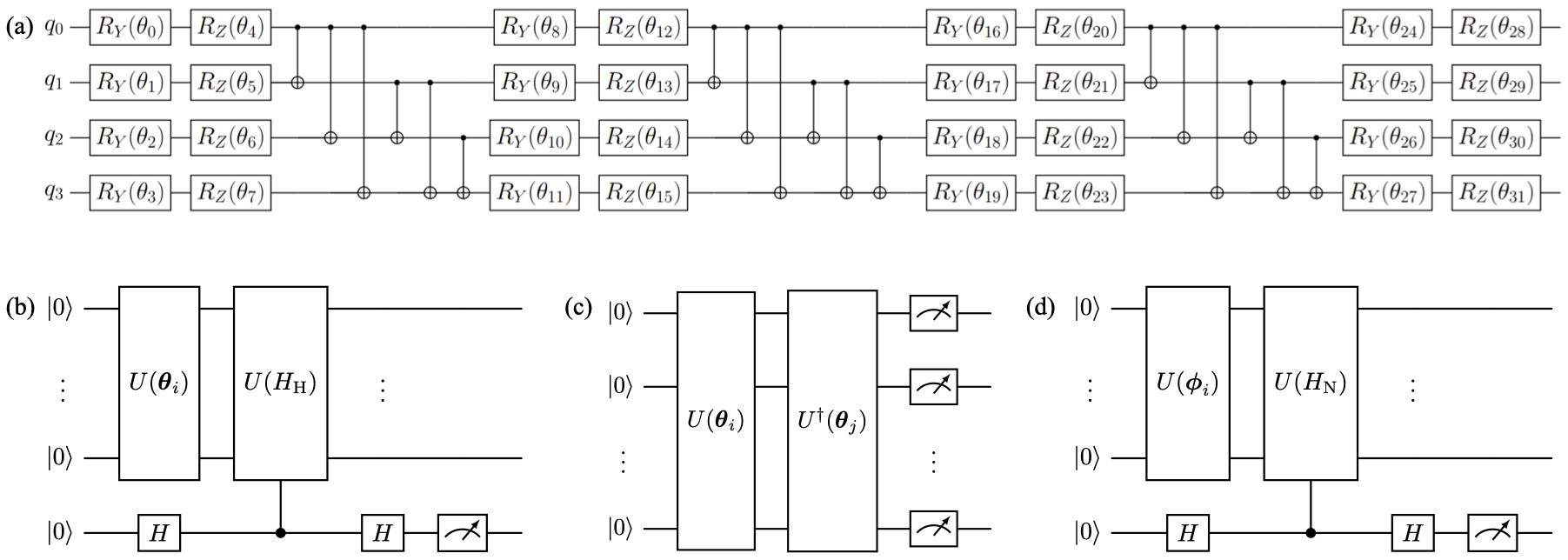}
\par\end{centering}
\caption{Quantum circuits for (a) the three-layer efficient SU(2) Ansatz, (b)
the Hadamard test for estimation of the expectation value of Pauli
word components of the Hermitian Hamiltonian $\text{Re}\left(\left\langle U\left(H_{\text{H}}\right)\right\rangle \right)$,
(c) the overlap between two states $\left\langle \psi\left(\mathbf{\theta}_{j}\right)\middle|\psi\left(\mathbf{\theta}_{i}\right)\right\rangle =\left\langle \bar{0}\left|U^{\dagger}\left(\mathbf{\theta}_{j}\right)U\left(\mathbf{\theta}_{i}\right)\right|\bar{0}\right\rangle $,
and (d) the Hadamard test for estimation of the expectation value
Pauli word components of the non-Hermitian Hamiltonian $\text{Re}\left(\left\langle U\left(H_{\text{N}}\right)\right\rangle \right)$.\label{fig:Circuits}}
\end{figure*}

To implement the aforementioned qDRIVE algorithm as a hybrid quantum-classical
approach, we begin by generating the Ansatz wavefunction as a quantum
circuit according to conventional VQE techniques,\cite{cao2019quantum,mcardle2020quantum,cerezo2021variational,Tilly2022Variational}
in which the Ansatz wavefunction consists of a series of unitary gates
parametrized by $m$ angles $\mathbf{\theta}_{i}$ for eigenstate
$i$, $\left|\psi\left(\mathbf{\theta}_{i}\right)\right\rangle =U\left(\mathbf{\theta}_{i}\right)\left|\bar{0}\right\rangle =U\left(\theta_{i_{m}}\right)U\left(\theta_{i_{m-1}}\right)\cdots U\left(\theta_{i_{1}}\right)\left|\bar{0}\right\rangle $,
where the form of the Ansatz is chosen to sufficiently balance expressibility
and circuit depth. To meet these requirements, here we employ the
three-layer efficient SU(2) Ansatz depicted in Fig.~\ref{fig:Circuits}(a)
with parameters initialized as pseudorandom angles in the domain $\theta_{i_{o}}\in\left[-\pi,\pi\right]$
for $o\in\left[1,m\right]$. 

Eigenstate identification for the Hermitian Hamiltonian $H_{\text{H}}$
is then performed via variational quantum deflation (VQD),\cite{higgott2019variational}
in which the ground state and successively identified eigenstates
are determined via VQE of an iteratively deflated Hamiltonian. Specifically,
the Hermitian Hamiltonian $H_{\text{H}}$ is represented as a weighted
sum of a set of unitary terms amenable to encoding on a $q$-qubit
quantum processor, here the $q$-qubit Pauli decomposition

\begin{align}
M & =\sum_{k_{1}=1}^{4}\sum_{k_{2}=1}^{4}\cdots\sum_{k_{q}=1}^{4}C_{k_{1},k_{2},\ldots,k_{q}}P_{k_{1},k_{2},\ldots,k_{q}}\label{eq:PauliDecomposition}\\
C_{k_{1},k_{2},\ldots,k_{q}} & =\text{Tr}\left(MP_{k_{1},k_{2},\ldots,k_{q}}\right)\\
P_{k_{1},k_{2},\ldots,k_{q}} & =s_{k_{1}}\otimes s_{k_{2}}\otimes\cdots\otimes s_{k_{q}}
\end{align}
where $P_{k_{1},k_{2},\ldots,k_{q}}$ denotes a Pauli word comprised
of Pauli gates $s_{k_{l}}\in\left\{ I,X,Y,Z\right\} $ for $j=1,2,\ldots,q$
\begin{equation}
I=\left[\begin{array}{cc}
1 & 0\\
0 & 1
\end{array}\right],\quad X=\text{\ensuremath{\left[\begin{array}{cc}
 0  &  1\\
 1  &  0 
\end{array}\right]}},\quad Y=\left[\begin{array}{cc}
0 & -\text{i}\\
\text{i} & 0
\end{array}\right],\quad Z=\left[\begin{array}{cc}
1 & 0\\
0 & -1
\end{array}\right].
\end{equation}
The objective function Eq.~(\ref{eq:VQDObjective}) is then estimated
on the quantum processor using the decomposition at values of $\mathbf{\theta}_{i}$
selected by a classical optimizer to converge towards the optimal
values associated with the Hermitian Hamiltonian eigenstates: The
classically expensive tasks of expectation value $\left\langle \psi\left(\mathbf{\theta}_{i}\right)\left|H_{\text{H}}\right|\psi\left(\mathbf{\theta}_{i}\right)\right\rangle $
and overlap $\left|\left\langle \psi\left(\mathbf{\theta}_{i}\right)\middle|\psi\left(\mathbf{\theta}_{j}\right)\right\rangle \right|^{2}$
estimation are performed on the quantum processor via the Hadamard
test shown in Fig.~\ref{fig:Circuits}(b) or a direct product of
qubit measurements in the basis of each Pauli word for the former
and the SWAP test,\cite{buhrman2001quantum,gottesman2001quantum}
the destructive SWAP test,\cite{garcia2013swap,cincio2018learning}
or the low-depth overlap method\cite{havlivcek2019supervised,higgott2019variational}
(shown in Fig.~\ref{fig:Circuits}(c)) for the latter; and the classically
efficient task of optimization is performed on the classical processor,
ensuring the choice of classical optimizer suits the degree of noise
associated with the quantum measurements and the choice of a penalty
parameter $c$ weighs sufficient deflation of the Hamiltonian with
ease of the optimization.\cite{kuroiwa2021penalty} According to these
considerations, here we employ a penalty parameter of $c=100$ with
$2^{9}$ maximum optimization iterations for all classical optimizations,
using COBYLA\cite{powell1994direct} (Constrained Optimization BY
Linear Approximation, initial variable change $P_{\text{beg}}=1$)
in the absence of noise and the NFT method\cite{Nakanishi2020NFT}
(the Nakanishi-Fujii-Todo method, maximum function evaluations
$f_{\max}=2^{11}$ and reset interval $R=32$) in the presence of
noise.

To converge the parameters associated with the Hermitian Hamiltonian
eigenstates $\mathbf{\theta}_{i}$ to the parameters associated with
the non-Hermitian Hamiltonian eigenstates $\mathbf{\phi}_{i}$, we
employ a hybrid quantum-classical approach to pseudovariance optimization.
We decompose the operators required for evaluation of the pseudovariance
objective function Eq.~(\ref{eq:PseudovarianceObjective}) $H_{\text{N}}$
and $H_{\text{N}}^{\dagger}H_{\text{N}}$ according to the Pauli decomposition
Eq.~(\ref{eq:PauliDecomposition}), employ the quantum processor to
estimate their expectation values (see example in Fig.~\ref{fig:Circuits}(d)),
and identify the optimal parameters $\mathbf{\phi}_{i}$ using a classical
optimizer suitable in both noiseless and noisy conditions, here Py-BOBYQA\cite{powell2009bobyqa,cartis2019improving}
(Bound Optimization BY Quadratic Approximation in Python, maximum
function evaluations $f_{\max}=2^{10}$, requested maximum pseudovariance
tolerance $f_{\text{tol}}=0.05$, maximum full retrials to reach tolerance
$u=3$, and initial trust region radius $R_{\text{beg}}=1$).

To reduce variance between instances of qDRIVE and \textcolor{black}{thereby increase the
robustness of the algorithm}, the aforementioned
procedure is executed as a batch of $\mathcal{B}$ runs (here $\mathcal{B}=8$).
Duplicate states identified within a single run are omitted where
the overlap between states is above a set tolerance, and the $i^{\text{th}}$
eigenstate of $H_{\text{N}}$ with the lowest value of the pseudovariance
across batches is considered the optimal representative of the state.
Nonresonant, diverging, and indifferent states\cite{Riss1993Calculation}
are then omitted through analysis of the eigenspectrum to produce
the terminal list of resonances.

\begin{figure}
\begin{centering}
\includegraphics[width=1\textwidth]{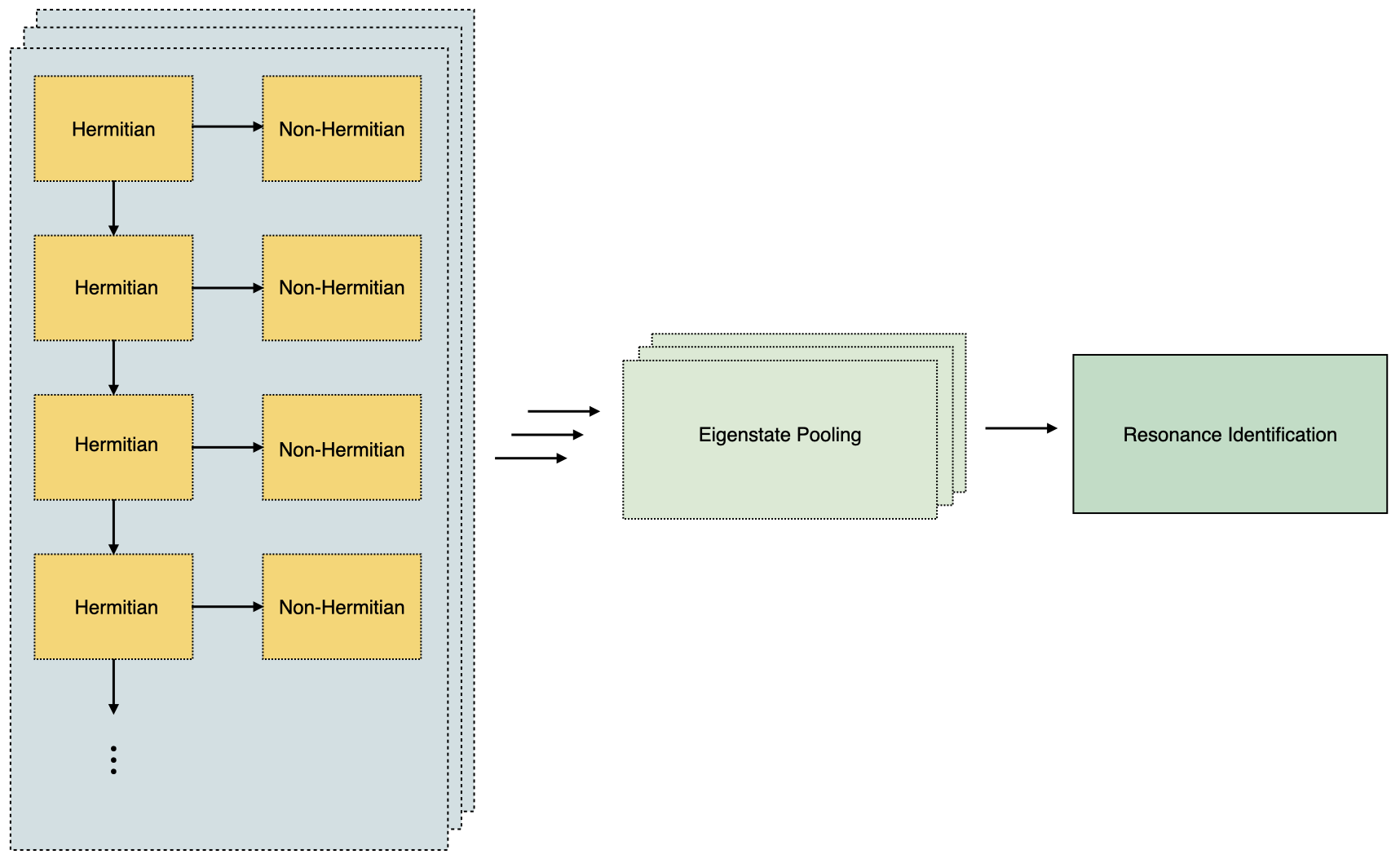}
\par\end{centering}
\caption{Schematic for implementation of qDRIVE with HTC resources. Each batch
(blue rectangles) begins with identification of the ground state of
the Hermitian Hamiltonian on an individual classical processor), which
spawns both identification of a corresponding eigenstate of the non-Hermitian
Hamiltonian on another classical processor and identification of the
next excited state on yet another classical processor (yellow rectangles).
For each batch, upon completion all resonances are pooled into a single
file (lime rectangles), and finally sorted among all batches to identify
the result for each resonance with the lowest pseudovariance (green
rectangle).\label{fig:HTCSchematic}}

\end{figure}

Given the asynchronicity of and minimal communication between the
aforementioned Hermitian and non-Hermitian Hamiltonian eigenstate
identification tasks, high-throughput computing (HTC)\cite{litzkow1987remote,litzkow1987condor,livny1997mechanisms,basney1999deploying,thain2005distributed,morgan2009high,fajardo2015much}
is used to minimize wall time to completion of the qDRIVE algorithm,
with underlying classical computing tasks executed in parallel according
to the directed acyclic graph illustrated in Fig.~\ref{fig:HTCSchematic}.
For each run, one classical processor identifies the ground state
of the Hermitian Hamiltonian $\psi\left(\mathbf{\theta}_{1}\right)$,
which initiates identification of the first eigenstate of the non-Hermitian
Hamiltonian $\psi\left(\mathbf{\phi}_{1}\right)$ on one processor
and identification of the first excited state of the Hermitian Hamiltonian
$\psi\left(\mathbf{\theta}_{2}\right)$ on another. Convergence of
the first excited state of the Hermitian Hamiltonian $\psi\left(\mathbf{\theta}_{2}\right)$
then spurs identification of the second excited state of the non-Hermitian
Hamiltonian $\psi\left(\mathbf{\phi}_{2}\right)$ and identification
of the second excited state of the Hermitian Hamiltonian $\psi\left(\mathbf{\theta}_{3}\right)$
and so on, incrementing $i\coloneqq i+1$ and proceeding until all
$N$ eigenstates of $H_{\text{N}}$ have been identified. The set
of $N$ eigenstates of $H_{\text{N}}$ for each run $\left\{ \psi\left(\mathbf{\phi}_{i}\right)\right\} $
are then communicated to a single processor per run to pool information
and finally on a single processor per batch to perform post-processing
and visualization. The opportunity for parallelism offered by the
directed acyclic graph may be paired with further parallelisms presented
by the structure of the system under study. For example, where both
the potential of interest $V_{0}$ and CAP $V_{\text{CAP}}$ are of
shared even parity, the eigenstates they support must also be of well-defined
parity such that even and odd states may be constructed independently;
where applicable, this parallelism (i) reduces the number of basis
states that must be represented on the quantum processor for a given
accuracy by half, which amounts to a reduction of the qubit count
by one, and (ii) enables classical tasks to be divided into two identical
copies of the same directed acyclic graph (one for each basis, odd
and even), which reduces the wall time by half. These wall time improvements
are obtained at the trade-off of a larger number of quantum circuit
evaluations and classical processors, respectively. Additionally,
this divided consideration of even and odd basis states simplifies
the optimization surface and ensures only basis states of suitable
parity contribute to the qDRIVE-optimized state.

\subsection{Near-Term Error Mitigation}

To facilitate the implementation of qDRIVE on near-term quantum computers,
we employ readout and gate error mitigation in conjunction with special
consideration of the identity Pauli word. 

\paragraph{Readout Error Mitigation}

Elementary readout error mitigation is performed via matrix inversion,
in which the true statistics of ancilla measurement are inferred from
noisy statistics\cite{dewes2012characterization,maciejewski2020mitigation,geller2020rigorous}
(see also advanced techniques\cite{cai2023quantum} such as twirled
readout error extinction\cite{karalekas2020quantum,van2022model}
to counteract noise-induced bias). Where the noisy expectation vector
$\vec{N}$ is related to the true expectation vector $\vec{T}$ by
\begin{equation}
\left[\begin{array}{c}
N_{0}\\
N_{1}
\end{array}\right]=\left[\begin{array}{c}
T_{0}\\
T_{1}
\end{array}\right]\left[\begin{array}{cc}
p_{00} & p_{01}\\
p_{10} & p_{11}
\end{array}\right]
\end{equation}
where the $0^{\text{th}}$ and the $1^{\text{st}}$ element of each
vector correspond to the probability the ancilla is measured in the
$\left|0\right\rangle $ and $\left|1\right\rangle $ state, respectively,
and where $p_{ij}$ is the probability the qubit is in true state
$i$ but measured in state $j$ in the presence of noise (determined
through system benchmarking\cite{lorenz2025systematic}); the true
expectation vector components are estimated as 
\begin{equation}
T_{0}=\frac{N_{0}-p_{10}}{p_{00}-p_{10}},\quad T_{1}=\frac{N_{1}-p_{01}}{p_{11}-p_{01}}
\end{equation}
according to solution of the system of equations given normalization
of the true expectation vector $T_{0}+T_{1}=1$ as expected of a classical
probability.

\paragraph{Gate Error Mitigation}

Gate error mitigation is performed via hybrid exponential-linear three-point
zero-noise extrapolation (ZNE).\cite{temme2017error,giurgica2020digital,miller2025universal}
According to standard ZNE, each circuit result $x_{\lambda}$ is extrapolated
to its zero-noise counterpart $\lambda=0$ based on the original result
of the circuit $\lambda=1$ and artificially increased-error iterations
of the circuit with $\left(\lambda-1\right)/2$ repetitions of each
gate and its inverse. We thus fit $x_{1}$, $x_{3}$, and $x_{5}$
to the exponential function 
\begin{equation}
x_{\lambda}\approx A+Be^{C\lambda}
\end{equation}
via solution of equations for the constants $A,B,C$ to yield the
zero-noise extrapolated circuit result\cite{miller2025universal}

\begin{equation}
x_{0}=x_{1}+\frac{x_{1}-x_{3}}{\beta^{2}+\beta},\quad\beta=\sqrt{\frac{x_{3}-x_{5}}{x_{1}-x_{3}}}.\label{eq:ExponentialFit}
\end{equation}
Importantly, the resulting zero-noise result $x_{0}$ is physically
realizable ($x_{0}\in\mathcal{R}$) if and only if $\left(x_{3}-x_{5}\right)/\left(x_{1}-x_{3}\right)>0$
(\emph{i.e.}, where $x_{\lambda}$ is monotonic for the three points
considered). We therefore employ physically motivated
linear approximations where monotonicity is violated to within statistical
significance as \textcolor{black}{detailed in the appendix}.

\paragraph{Identity Pauli Word}

Additionally, we make the simple recognition that the expectation
value of the identity Pauli word $I^{\otimes q}$ is analytically known
to be unity

\begin{equation}
\left\langle \psi\left|I^{\otimes q}\right|\psi\right\rangle =\left\langle \psi\middle|\psi\right\rangle =1
\end{equation}
for all normalized states $\psi$, such that the expectation value
need not be estimated on the quantum processor. This small point ensures
accurate evaluation of the term free of the vicissitudes of readout
or gate error, of particular importance for the investigation of chemical
Hamiltonians (and associated pseudovariances) for which its associated
weight in the Pauli decomposition Eq.~(\ref{eq:PauliDecomposition})
is relatively large.

\subsection{Model System}

\begin{figure}
\begin{centering}
\includegraphics[width=0.75\textwidth]{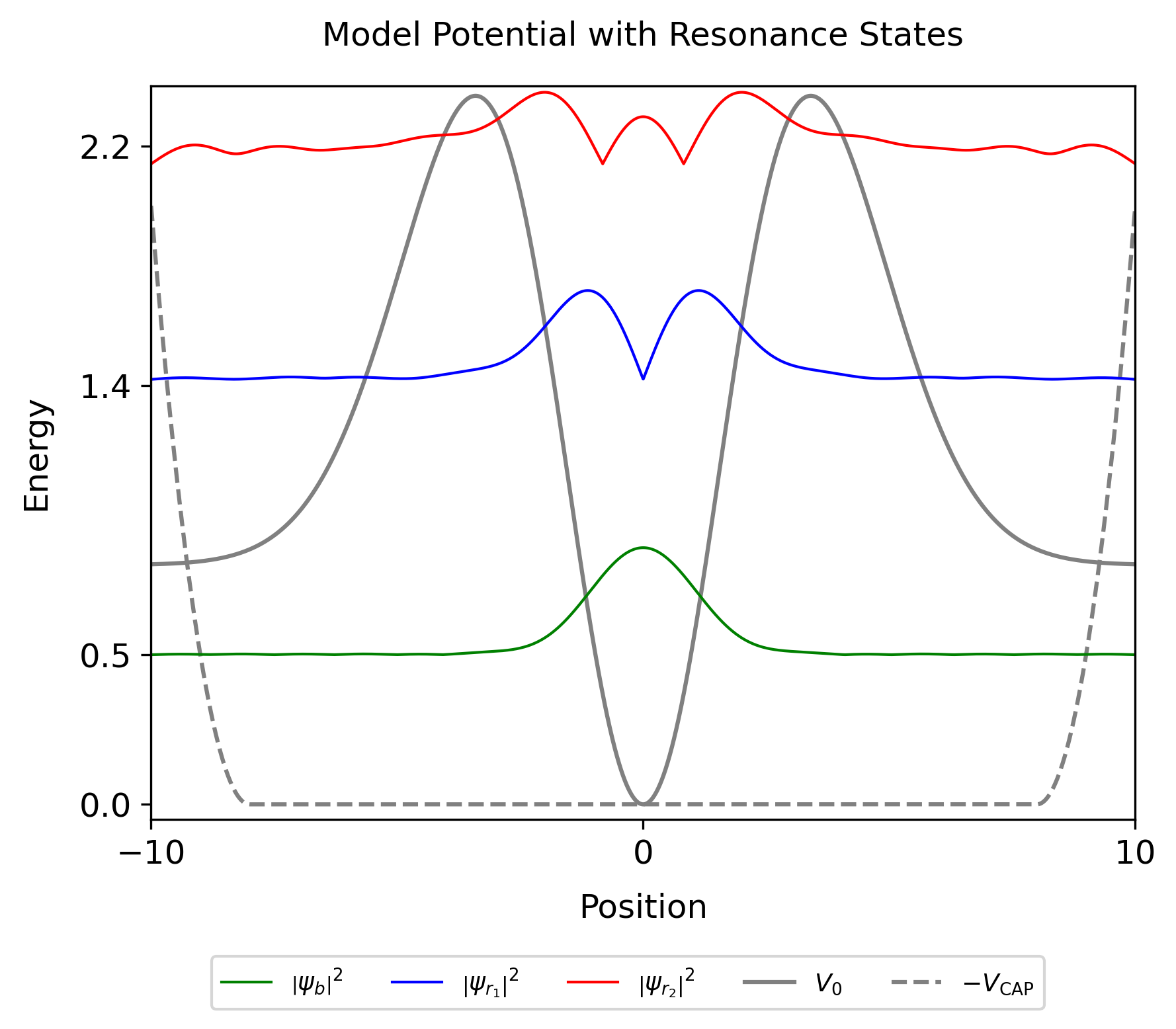}
\par\end{centering}
\caption{Benchmark potential energy surface $V_{0}$ (solid gray line), which
supports a bound state $\psi_{b}$ and two resonances $\psi_{r_{1}}$
and $\psi_{r_{2}}$ (solid green, blue, and red lines, probability
density shown), here termed the bound state, first resonance, and
second resonance, respectively, shown as computed via the \textcolor{black}{established classical CAP method} of exact diagonalization
where purely outgoing boundary conditions are imposed by a complex
absorbing potential $V_{\text{CAP}}$ (dashed gray line, shown as
$-V_{\text{CAP}}$). Each state's probability density is vertically
shifted by the value of the real part of its energy $\text{Re}\left(E\right)$.\label{fig:Benchmark-potential-energy}}

\end{figure}

To demonstrate the power of the qDRIVE algorithm, we present the method
as applied to a benchmark potential designed to model resonances associated
with diatomic predissociation and molecular scattering collisions\cite{moiseyev1978resonance}
\begin{equation}
V_{0}\left(x\right)=\left(\frac{1}{2}x^{2}-J\right)\text{e}^{-\lambda x^{2}}+J\label{eq:BenchmarkPotential}
\end{equation}
where $\lambda=0.1$ and $J=0.8$ (arbitrary units by convention),
with purely outgoing boundary conditions imposed via a quadratic CAP
\begin{equation}
V_{\text{CAP}}\left(x\right)=\begin{cases}
0 & \left|x\right|\le x_{0}\\
-\frac{1}{2}\left(\left|x\right|-x_{0}\right)^{2} & \left|x\right|>x_{0}
\end{cases}
\end{equation}
where $x_{0}=8\text{ au}$, as illustrated in Fig.~\ref{fig:Benchmark-potential-energy}.
To aid direct comparison to literature results, we specifically examine
the resonance near $\text{Re}\left(E_{r_{2}}\right)=2.13$ investigated
by Bian et al.~with the direct measurement method.\cite{bian2019quantum}
To demonstrate the breadth of the method, we additionally employ qDRIVE
to identify the bound state near $E_{b}=0.502$
and the narrower resonance near $\text{Re}\left(E_{r_{1}}\right)=1.42$. Simulations shown
here are performed for a Hamiltonian implemented in position-space
representation in the domain $x\in\left[-x_{\max},x_{\max}\right]$ with
$x_{\max}=10$ and $2^{12}$ gridpoints, a kinetic energy operator
computed according to the second-order finite differencing gradient,
and a basis set comprising of $2^{q}$ sinusoidal basis states of
length $L=2x_{\max}$ 
\begin{equation}
\phi_{k}\left(x\right)=\sqrt{\frac{2}{L}}\sin\left(\frac{w\pi}{L}\left(x-\frac{L}{2}\right)\right)
\end{equation}
of either even $w=1,2,3,\ldots$ or odd $w=2,4,6,...$ symmetry.

\subsection{Quantum Simulation \textcolor{black}{and Experimental Tests}}

All codes for quantum simulation and experimental tests have been made available publicly
under an open-source license from Github\cite{qdrivegithub} using
IBM Qiskit.\cite{javadi2024quantum} The statevector simulator is
used to examine qDRIVE in the absence of noise; the Aer simulator,
in the presence of only statistical shot noise; and a custom simulator,
in the presence of statistical shot noise, readout noise, and gate
noise. \textcolor{black}{IBM Falcon r4T processors are used to examine the efficacy of
 real quantum computers for the most quantum-processor-intensive stage of qDRIVE (namely, pseudovariance minimization).}
 
\textcolor{black}{In simulations, where} statistical shot noise is incorporated, $n=10^5$
shots are taken per quantum circuit. An additional order of magnitude
of shots is employed for final comparison of results between runs.
Where readout and gate error are included, the noise model for thermal
relaxation at time $t$ incorporates both the generalized amplitude
damping channel Kraus matrices
\begin{align}
K_{0} & =\sqrt{1-p}\left[\begin{array}{cc}
1 & 0\\
0 & \sqrt{1-\gamma_{1}}
\end{array}\right],\,K_{1}=\sqrt{1-p}\left[\begin{array}{cc}
0 & \sqrt{\gamma_{1}}\\
0 & 0
\end{array}\right],\\
K_{2} & =\sqrt{p}\left[\begin{array}{cc}
1 & 0\\
0 & \sqrt{1-\gamma_{1}}
\end{array}\right],\,K_{3}=\sqrt{p}\left[\begin{array}{cc}
0 & \sqrt{\gamma_{1}}\\
0 & 0
\end{array}\right],\\
\gamma_{1} & =1-\text{e}^{-t/T_{1}},
\end{align}
where $p$ is the equilibrium excited state population and $T_{1}$
is the generalized amplitude damping relaxation time, and the phase
damping model Kraus matrices
\begin{align}
K_{0} & =\left[\begin{array}{cc}
1 & 0\\
0 & \sqrt{1-\gamma_{2}}
\end{array}\right],\quad K_{1}=\left[\begin{array}{cc}
0 & 0\\
0 & \sqrt{\gamma_{2}}
\end{array}\right]\\
\gamma_{2} & =1-\text{e}^{-t/T_{2}},
\end{align}
where $T_{2}$ is the phase damping relaxation time. The local one-
and two-qubit depolarizing error is then described by the error channel
\begin{equation}
E\left(\rho\right)=\left(1-p_{d}\right)\rho+p_{d}\text{Tr}\left[\rho\right]\frac{I}{2^{d}}
\end{equation}
for $d$-qubit gate error probability $p_{d}$. In the custom simulator
model of IBM Torino, qubit and processor specifications are hard-coded
to approximate the contemporary properties of the quantum processor,
including system basis gates and qubit-specific one- and
two-qubit gate errors, readout errors, $T_{1}/T_{2}$ times, and gate
times. To reduce simulation cost, only five qubits of the full IBM
Torino system are simulated to accommodate a maximal four-qubit Ansatz
and a single ancilla. Quantum processors with specifications beyond
those of IBM Torino are simulated using the same custom simulator
with a gate noise reduction factor applied to the one- and two-qubit
gate error probabilities $p_{1}$ and $p_{2}$ and a qubit longevity
factor that represents the order of magnitude of the $T_{1}$ relaxation
time where the leading digit of $T_1$ and the ratio between the $T_{1}$ and $T_{2}$ relaxation
times is held fixed. For example, where the IBM Torino relaxation
times $T_{1}$ and $T_{2}$ are approximated as $7\text{0  \textmu s}$ and
$50\text{  \textmu s}$, respectively, a qubit longevity factor of $100\text{  \textmu s}$
implies relaxation times $T_{1}$ and $T_{2}$ of $700\text{  \textmu s}$
and $500\text{  \textmu s}$, respectively. Infinite qubit longevity factors
correspond to the absence of thermal relaxation. 

\textcolor{black}{In experimental tests on real quantum processors, the most computationally intensive stage of qDRIVE for the quantum processor, namely, pseudovariance minimization, is tested on five-qubit IBM Falcon r4T quantum processors Quito, Belem, and Lima. Device specifications, including qubit relaxation times and gate error rates, are provided for reference in Table~\ref{tab:IBMRealSpecs}. To isolate error due to pseudovariance minimization from that of variational quantum deflation (VQD), Ansatz parameters are randomly initialized in the domain $\theta_{i_{o}}\in\left[-\pi,\pi\right]$ where $o\in\left[1,m\right]$. A two-local circuit with $R_x$ and $R_y$ rotation gates is employed with full entanglement between qubits provided by CNOT gate layers in order to balance Ansatz expressibility with quantum processor cost where applicable. Energies are computed for one- and two-qubit Hamiltonians represented in a sinusoidal basis using a shot number of $n=10^4$ on the quantum processor and COBYLA\cite{powell1994direct} (tolerance $10^{-3}$, maximum iteration number $15$) on the classical processor. Given limited quantum processing resources, real quantum computer tests are automatically executed on the least occupied quantum processor and in such tests no error mitigation methods are employed.}

Results are reported in terms of energies computed at the level of
quantum simulation \textcolor{black}{or quantum computation} used and eigenstate probability densities are visualized where relevant
with the statevector simulator. 
\textcolor{black}{To evaluate the success of qDRIVE, qDRIVE's accuracy, robustness, and 
computational cost are compared to that of two established classical resonance identification techniques: 
(1) the established classical CAP resonance identification method
in which resonances are determined via exact diagonalization of
$H_\text{N}=H_\text{H}+\text{i}V_\text{CAP}$,\cite{kosloff1984dynamical,jolicard1985optical,kosloff1986absorbing,neuhasuer1989time,Muga2004Complex} and (2) the  established classical complex scaling method 
in which resonances are determined via $\theta$ trajectories of the complex-scaled Hamiltonian.\cite{aguilar1971class,balslev1971spectral,bian2019quantum} 
Classical CAP method results are presented for identical basis sets to that of qDRIVE for direct comparison, and classical complex scaling method results
are presented for a basis set of five orthogonalized Gaussian basis states $\chi_k(\alpha)=\exp(-\alpha_k x^2)$ with $\alpha_k=0.65\cdot 0.45^k$ where $k=0,1,2,3,4$
for comparison to the literature solution of ref.~\citenum{bian2019quantum}.}

\begin{table*}

\color{black}

\begin{centering}
\begin{tabular}{cccccccccc}
\toprule
\multirow{2}{*}{Device} & \multirow{2}{*}{Qubit} & \multirow{2}{*}{$T_1$} & \multirow{2}{*}{$T_2$} & \multicolumn{5}{c}{Error}  & \multirow{2}{*}{Gate Time} \tabularnewline 
\cmidrule{5-9}
 &  &  &  & Readout & $I$ & $\sqrt{X}$ & $X$  & CNOT &   \tabularnewline
\midrule
\midrule
\multirow{9}{*}{\makecell{Quito \\ (1.1.45)}} & 0 & 112.5 & 139.23 & 3.340 & 2.437 & 2.437 & 2.437 & 0$\_$1:0.03884 & 0$\_$1:234.667 \tabularnewline
%\cline{2-10}
& \multirow{3}{*}{1} & \multirow{3}{*}{106.41} & \multirow{3}{*}{101.49} & \multirow{3}{*}{3.640} & \multirow{3}{*}{2.726} & \multirow{3}{*}{2.726} & \multirow{3}{*}{2.726} & 1$\_$3:0.00958 & 1$\_$3:334.222  \tabularnewline
& & & & & & & & 1$\_$2:0.01122 & 1$\_$2:298.667 \tabularnewline
& & & & & & & & 1$\_$0:0.03884 & 1$\_$0:270.222 \tabularnewline
%\cline{2-10}
& 2 & 63.66 & 37.37 & 7.640 & 3.474 & 3.474 & 3.474 & 2$\_$1:0.01122 & 2$\_$1:263.111  \tabularnewline
%\cline{2-10}
& \multirow{2}{*}{3} & \multirow{2}{*}{58.45} & \multirow{2}{*}{18.14} & \multirow{2}{*}{3.920} & \multirow{2}{*}{2.713} & \multirow{2}{*}{2.713} & \multirow{2}{*}{2.713} & 3$\_$4:0.01469 & 3$\_$4:277.333  \tabularnewline
& & & & & & & & 3$\_$1:0.00958 & 3$\_$1:369.778 \tabularnewline
%\cline{2-10}
& 4 & 70.44 & 131.29 & 4.05 & 2.767 & 2.767 & 2.767 & 4$\_$3:0.01469 & 4$\_$3:312.889  \tabularnewline
\midrule
%----------------------
\multirow{9}{*}{\makecell{Belem \\ (1.2.12)}} & 0 & 116.9 & 135.85 & 2.740 & 2.104 & 2.104 & 2.104 & 0$\_$1:0.01478 & 0$\_$1:810.667 \tabularnewline
%\cline{2-10}
& \multirow{3}{*}{1} & \multirow{3}{*}{93.12} & \multirow{3}{*}{92.11} & \multirow{3}{*}{1.960} & \multirow{3}{*}{4.920} & \multirow{3}{*}{4.920} & \multirow{3}{*}{4.920} & 1$\_$3:0.00669 & 1$\_$3:440.889  \tabularnewline
& & & & & & & & 1$\_$2:0.00586 & 1$\_$2:419.556 \tabularnewline
& & & & & & & & 1$\_$0:0.01478 & 1$\_$0:775.111 \tabularnewline
%\cline{2-10}
& 2 & 58.22 & 47.91 & 1.65 & 1.976 & 1.976 & 1.976 & 2$\_$1:0.00586 & 2$\_$1:384.00  \tabularnewline
%\cline{2-10}
& \multirow{2}{*}{3} & \multirow{2}{*}{88.4} & \multirow{2}{*}{94} & \multirow{2}{*}{2.63} & \multirow{2}{*}{2.721} & \multirow{2}{*}{2.721} & \multirow{2}{*}{2.721} & 3$\_$4:0.00908 & 3$\_$4:526.222  \tabularnewline
& & & & & & & & 3$\_$1:0.00669 & 3$\_$1:405.333 \tabularnewline
%\cline{2-10}
& 4 & 58.15 & 45.07 & 1.73 & 5.656 & 5.656 & 5.656 & 4$\_$3:0.00908 & 4$\_$3:490.667  \tabularnewline
\midrule
%----------------------
\multirow{9}{*}{\makecell{Lima \\ (1.0.52)}} & 0 & 113.67 & 160.08 & 1.96 & 7.567 & 7.567 & 7.567 & 0$\_$1:0.00662 & 0$\_$1:305.778 \tabularnewline
%\cline{2-10}
& \multirow{3}{*}{1} & \multirow{3}{*}{93.43} & \multirow{3}{*}{131.71} & \multirow{3}{*}{1.740} & \multirow{3}{*}{5.44} & \multirow{3}{*}{5.44} & \multirow{3}{*}{5.44} & 1$\_$0:0.00662 & 1$\_$0:341.333  \tabularnewline
& & & & & & & & 1$\_$3:0.01152 & 1$\_$3:497.778 \tabularnewline
& & & & & & & & 1$\_$2:0.00695 & 1$\_$2:334.222 \tabularnewline
%\cline{2-10}
& 2 & 69.01 & 99.69 & 2.06 & 7.334 & 7.334 & 7.334 & 2$\_$1:0.00695 & 2$\_$1:298.667  \tabularnewline
%\cline{2-10}
& \multirow{2}{*}{3} & \multirow{2}{*}{59.08} & \multirow{2}{*}{91.58} & \multirow{2}{*}{3.85} & \multirow{2}{*}{2.367} & \multirow{2}{*}{2.367} & \multirow{2}{*}{2.367} & 3$\_$4:0.01476 & 3$\_$4:519.111  \tabularnewline
& & & & & & & & 3$\_$1:0.01152 & 3$\_$1:462.222 \tabularnewline
%\cline{2-10}
& 4 & 16.13 & 16.17 & 4.43 & 6.514 & 6.514 & 6.514 & 4$\_$3:0.01476 & 4$\_$3:483.556  \tabularnewline
\bottomrule

\end{tabular}
\end{centering}

\caption{\textcolor{black}{Specifications of IBM Falcon r4T quantum processor devices (version in parenthesis) employed in experiments, as of September 14, 2023, from IBM Qiskit\cite{javadi2024quantum}. Values are reported in microseconds for relaxation times $T_1$ and $T_2$, nanoseconds for two-qubit gate times, and with implied multiplicative factors of $10^{-2}$, $10^{-4}$, $10^{-4}$, $10^{-4}$, and $10^{-3}$ for readout assignment, $I$, $\sqrt{X}$, $X$, and CNOT error rates, respectively.\label{tab:IBMRealSpecs}}}
\end{table*}

\subsection{Results}

\begin{table*}
\begin{centering}
\begin{tabular}{ccccc}
\toprule 
$q$ & Sim. & $E_{b}$ & $E_{r_{1}}$ & $E_{r_{2}}$\tabularnewline
\midrule
\midrule 
$2$ & \textcolor{black}{Classical CAP Method} & $0.623-2.63\cdot10^{-3}\text{i}$ & $1.61-4.15\cdot10^{-2}\text{i}$ & $2.36-5.83\cdot10^{-3}\text{i}$\tabularnewline
 & Statevector & $0.623-2.78\cdot10^{-3}\text{i}$ & $1.61-4.26\cdot10^{-2}\text{i}$ & $2.34-1.22\cdot10^{-2}\text{i}$\tabularnewline
 & Aer & $0.619-2.97\cdot10^{-3}\text{i}$ & $1.61-3.97\cdot10^{-2}\text{i}$ & $2.35-7.13\cdot10^{-3}\text{i}$\tabularnewline
 & Torino & $0.685-2.37\cdot10^{-3}\text{i}$ & $1.61-5.62\cdot10^{-2}\text{i}$ & $2.20-1.29\cdot10^{-2}\text{i}$\tabularnewline
\midrule 
$3$ & \textcolor{black}{Classical CAP Method} & $0.505-2.02\cdot10^{-5}\text{i}$ & $1.43-1.61\cdot10^{-4}\text{i}$ & $2.15-2.04\cdot10^{-2}\text{i}$\tabularnewline
 & Statevector & $0.504-2.48\cdot10^{-5}\text{i}$ & $1.43-1.44\cdot10^{-4}\text{i}$ & $2.14-1.00\cdot10^{-2}\text{i}$\tabularnewline
 & Aer & $0.505+2.43\cdot10^{-4}\text{i}$ & $1.44+2.60\cdot10^{-4}\text{i}$ & $2.09-2.64\cdot10^{-2}\text{i}$\tabularnewline
 & Torino & $0.562+2.22\cdot10^{-3}\text{i}$ & $1.92-8.85\cdot10^{-2}\text{i}$ & $2.18-1.09\cdot10^{-2}\text{i}$\tabularnewline
\midrule 
$4$ & \textcolor{black}{Classical CAP Method} & $0.502-9.98\cdot10^{-11}\text{i}$ & $1.42-3.60\cdot10^{-5}\text{i}$ & $2.12-1.18\cdot10^{-2}\text{i}$\tabularnewline
 & Statevector & $0.502-1.23\cdot10^{-7}\text{i}$ & $1.42-4.03\cdot10^{-5}\text{i}$ & $2.13-4.63\cdot10^{-4}\text{i}$\tabularnewline
\bottomrule
\end{tabular}\medskip{}
\par\end{centering}
\begin{centering}
\begin{tabular}{ccr@{\extracolsep{0pt}.}lcc}
\toprule 
$q$ & Sim. & \multicolumn{2}{c}{$\mathcal{E}_{b}$} & $\mathcal{E}_{r_{1}}$ & $\mathcal{E}_{r_{2}}$\tabularnewline
\midrule
\midrule 
$2$ & Statevector & 0&02\% & 0.07\% & 0.89\%\tabularnewline
 & Aer & 0&64\% & 0.11\% & 0.43\%\tabularnewline
 & Torino & 9&95\% & 0.91\% & 6.79\%\tabularnewline
\midrule 
$3$ & Statevector & 0&20\% & 0.0012\% & 0.67\%\tabularnewline
 & Aer & 0&05\% & 0.70\% & 2.80\%\tabularnewline
 & Torino & 11&30\% & 34.82\% & 1.46\%\tabularnewline
\midrule 
$4$ & Statevector & 0&000024\% & 0.00030\% & 0.71\%\tabularnewline
\bottomrule
\end{tabular}
\par\end{centering}
\caption{(a) Bound and resonance state energies $E_{b}$, $E_{r_{1}}$, and
$E_{r_{2}}$ computed according to qDRIVE for a $q$-qubit Ansatz
via the quantum simulators (Sim.) Statevector, Aer, and IBM Torino,
as compared to the result of \textcolor{black}{the classical CAP method} according to (b)
percent relative error $\mathcal{E}=\left|z^{\prime}-z\right|/\left|z\right|$.\label{tab:StatevectorAerTorinoEnergies}}
\end{table*}

The interlaced quantum computing/HTC qDRIVE approach was found to \textcolor{black}{accurately}
identify all resonance energies considered for the benchmark resonance
model of molecular predissociation Eq.~(\ref{eq:BenchmarkPotential}) 
\textcolor{black}{as compared to established classical resonance identification techniques.}
As shown in Table~\ref{tab:StatevectorAerTorinoEnergies}, the error
of the bound-state energy $E_{b}$, the first-resonance energy $E_{r_{1}}$,
and the second-resonance energy $E_{r_{2}}$ relative to \textcolor{black}{the established classical CAP method} 
results was found to remain below $1\%$ in all statevector simulations
for two- to four-qubit Ans{\"a}tze. Additionally, the relative error was
observed to be as low as $O\left(10^{-5}\%\right)$ for the highest
qubit-number statevector simulation considered, indicative of the
success of the method in the absence of shot noise and readout and
gate error. In experiments with the Aer simulator that included shot
noise, the relative error remained below $1\%$ in all simulations
with the exception of a relative error of $2.8\%$ for $E_{r_{2}}$
where a three-qubit Ansatz was employed, which was also associated
with larger componentwise error of $\text{Im}\left(E_{b}\right)$
and $\text{Im}\left(E_{r_{1}}\right)$. As expected, in custom simulations
of IBM Torino that included readout and gate error, the relative error
was larger but still in the vicinity of \textcolor{black}{classical CAP method} results,
with an error as low as $0.91\%$ for $E_{r_{1}}$ with a two-qubit
Ansatz and as high as $35\%$ for the same energy $E_{r_{1}}$ with
a three-qubit Ansatz, which suggests qDRIVE may be already implemented
as specified with existing quantum computing systems where the error
tolerance is acceptable. \textcolor{black}{qDRIVE results for the second resonance energy
$E_{r_2}$ were also found to be accurate relative to the literature classical complex
scaling method result from ref.~\citenum{bian2019quantum} of $E=2.1265-0.0203\text{i}$. As expected,
the accuracy of qDRIVE relative to the established classical complex scaling resonance identification technique was found
to increase as the basis set size increased, with 
a relative error of $10.0\%$, $10.5\%$, and $3.5\%$ for two-qubit statevector, Aer, and Torino, respectively; 
$0.8\%$, $1.7\%$, and $2.6\%$ for three-qubit statevector, Aer, and Torino, respectively; and $0.9\%$ for four-qubit statevector. Note the qDRIVE simulations were also found to be highly robust relative to established classical resonance identification techniques, as the bound and resonance states were
identified in $100\%$ of the quantum simulations for the number of batch runs considered, in full agreement
with the robustness of classical CAP and classical complex scaling methods.}

\textcolor{black}{Results of the real quantum computer experiments, shown in Table~\ref{tab:RealQC}, indicate that pseudovariance minimization successfully yielded complex energies near eigenvalues of the Hamiltonian. Both eigenvalues of the one-qubit Hamiltonian were identified, with a relative error of $17.0\%$ and $5.1\%$ relative to the classical CAP method result. The two complex energies of the two-qubit Hamiltonian with a real part closest to that of the scattering $r_1$ and $r_2$ resonances observed for larger basis sets were likewise measured with $30.5\%$ and $8.9\%$ relative error relative to the classical CAP method of exact diagonalization of the same Hamiltonian, respectively. As expected, the resonance energies were therefore within the vicinity of but with a higher degree of relative error than statevector, error-mitigated Aer, and error-mitigated simulated IBM Torino resonance energies for the system examined in Table~\ref{tab:StatevectorAerTorinoEnergies}.}

\begin{figure*}
\begin{centering}
\includegraphics[width=1\textwidth]{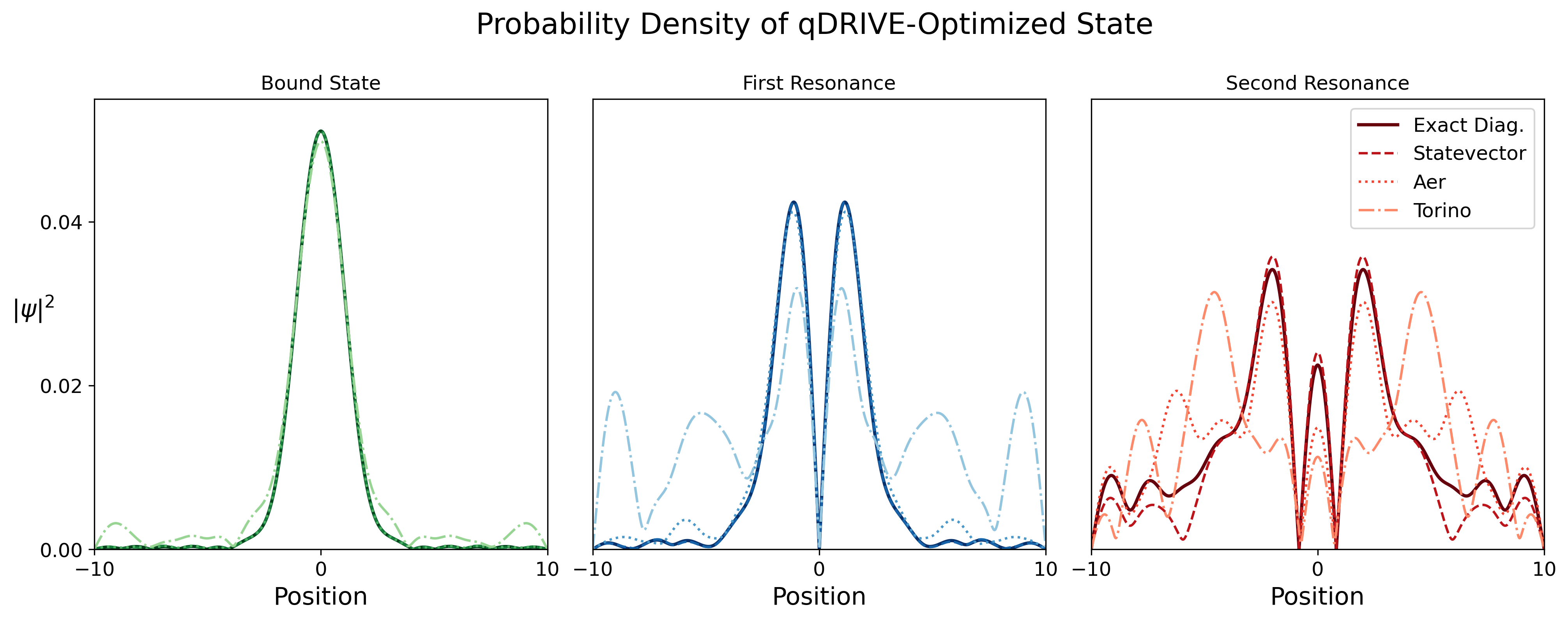}
\par\end{centering}
\caption{Probability \textcolor{black}{density} $\left|\psi\right|^{2}$ visualized in position
space corresponding to the qDRIVE-optimized three-qubit Ansatz for
the statevector (dashed lines), Aer (dotted lines), and Torino (dashed-dotted
lines) simulators as compared to \textcolor{black}{classical CAP method/}exact diagonalization (solid lines)
for the (left) bound state (green lines), (center) first resonance
(blue lines), and (right) second resonance (red lines).\label{fig:Probability-density}}
\end{figure*}

\begin{table}

\begin{centering}

\color{black}

\begin{tabular}{cccr@{\extracolsep{0pt}.}l}
\toprule 
$q$ & Classical CAP Method & IBM Falcon r4T & \multicolumn{2}{c}{$\mathcal{E}$}\tabularnewline
\midrule
\midrule 
\multirow{2}{*}{1} & $0.55+0.0000\text{i}$ & $0.54+0.093\text{i}$ & \multicolumn{2}{c}{$17.0\%$}\tabularnewline
%\cmidrule{2-5} 
 & $2.22+0.0000\text{i}$ & $2.11+0.0248\text{i}$ & \multicolumn{2}{c}{$5.1\%$}\tabularnewline
\midrule 
\multirow{2}{*}{2} & $1.56+0.0698\text{i}$ & $1.39-0.3745\text{i}$ & \multicolumn{2}{c}{$30.5\%$}\tabularnewline
%\cmidrule{2-5} 
 & $2.26-0.0034\text{i}$ & $2.06-0.007\text{i}$ & \multicolumn{2}{c}{$8.9\%$}\tabularnewline
\bottomrule
\end{tabular}
\end{centering}
\caption{\textcolor{black}{Energies of selected states computed via qDRIVE pseudovariance minimization of a randomly initiated $q$-qubit Ansatz on IBM Falcon r4T Quito, Lima, and Belem processors, with percent relative error $\mathcal{E}$ calculated relative to the classical CAP method of exact diagonalization of the corresponding Hamiltonian. Note the two-qubit classical CAP method values are distinct from those of Table~\ref{tab:StatevectorAerTorinoEnergies} as a  combined even/odd basis set is employed.\label{tab:RealQC}}}
\end{table}

qDRIVE was also found to successfully produce probability densities
consistent with \textcolor{black}{classical CAP method} results for all resonances considered,
as shown in Fig.~\ref{fig:Probability-density} for a three-qubit
Ansatz. Probability densities from statevector simulations closely
agreed with \textcolor{black}{classical CAP method} results for the bound state and
first resonance, with a higher discrepancy in the second resonance
that may be attributable to the underlying deflation procedure given
the high-lying nature of its corresponding Hermitian Hamiltonian eigenstate.
Probability densities from Aer simulations were found to be accurate
in the internal region of the potential, with a lower degree of accuracy
than statevector, as expected due to the additional consideration
of statistical noise. Consistent with the impact of further readout
and gate noise, the greatest discrepancy between the qDRIVE-optimized
and \textcolor{black}{classical-CAP-method} probability densities was observed for the
custom simulations of IBM Torino; nonetheless, overall trends of localization
of the bound state and resonances in the potential's internal region
and reduction in the outer region were preserved.

\begin{figure*}
\begin{centering}
\includegraphics[width=1\textwidth]{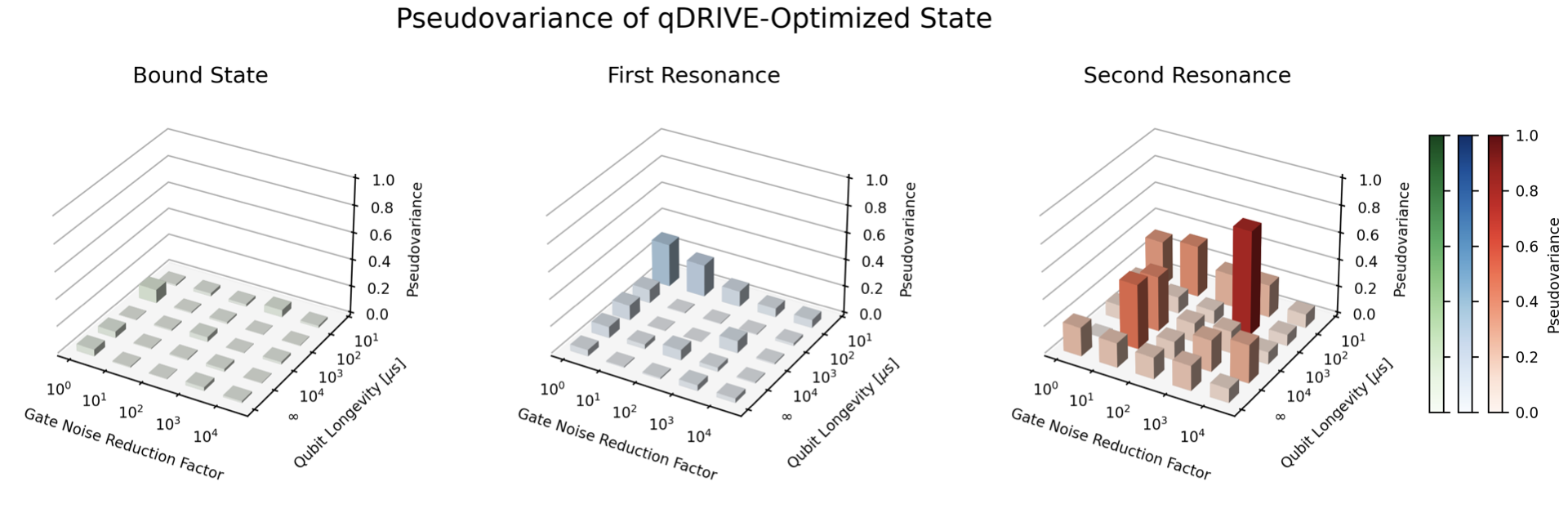}
\par\end{centering}
\caption{Pseudovariance Eq.~(\ref{eq:Pseudovariance}) of the (left) bound
state, (center) first resonance, and (right) second resonance as determined
by qDRIVE with a three-qubit Ansatz as a function of the gate noise
reduction and qubit longevity factors.\label{fig:Pseudovariance}}
\end{figure*}

\begin{figure*}
\begin{centering}
\includegraphics[width=1\textwidth]{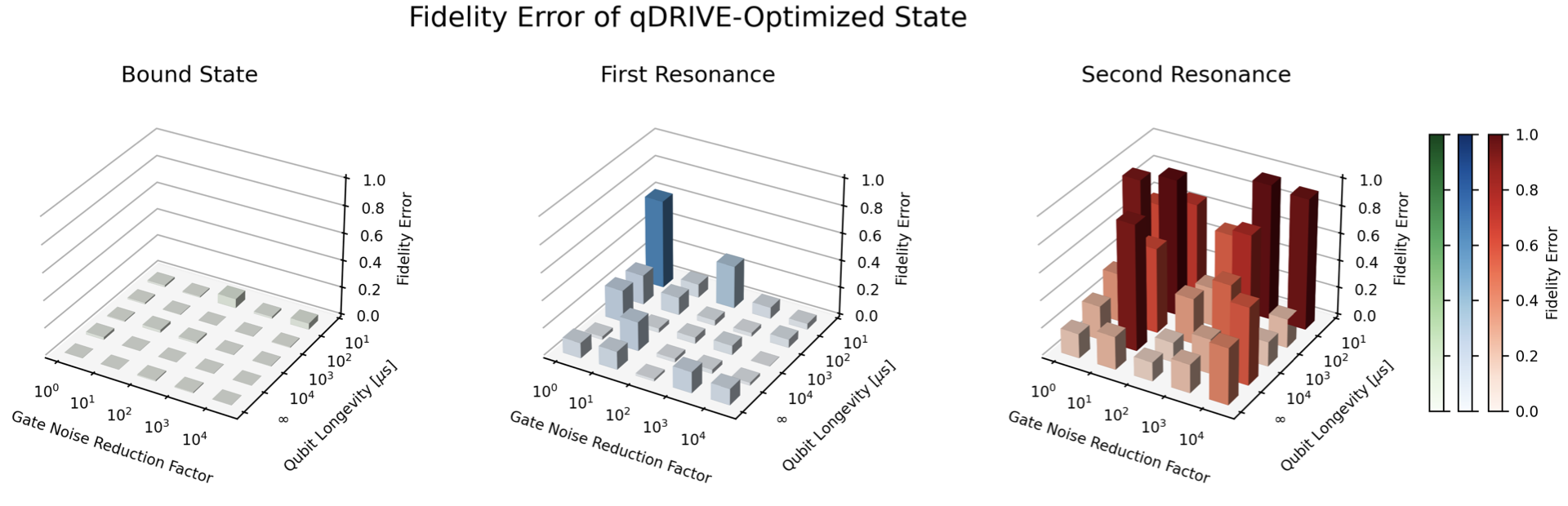}
\par\end{centering}
\caption{Fidelity error of the simulated qDRIVE-optimized (left) bound state,
(center) first resonance, and (right) second resonance $\psi_{\text{Sim}}$
as compared to the \textcolor{black}{classical CAP method/}exact diagonalization result $\psi_{\text{Exact}}$,
$1-\left|\left\langle \psi_{\text{Sim}}\middle|\psi_{\text{Exact}}\right\rangle \right|^{2}$,
for increasing gate noise reduction and qubit longevity factors.\label{fig:FidelityError}}
\end{figure*}

Increased gate noise reduction and qubit longevity factors were associated
with lower pseudovariance and fidelity error of the qDRIVE-optimized
resonances, as shown in Fig.~\ref{fig:Pseudovariance} and Fig.~\ref{fig:FidelityError}.
Both the pseudovariance and the fidelity error were found to approach
their optimal zero value of zero as the gate noise reduction factor increased
from $10^{0}$ to $10^{4}$ and the qubit longevity increased from
$10^{1}$ to $\infty\text{  \textmu s}$, with fluctuations in keeping with
the statistical nature of the simulations. An increase in the maximal
value of the pseudovariance was also observed for states corresponding
to higher-lying eigenstates of the Hermitian Hamiltonian in keeping
with errors introduced by deflation.
 
\begin{figure*}
\begin{centering}
\includegraphics[width=1\textwidth]{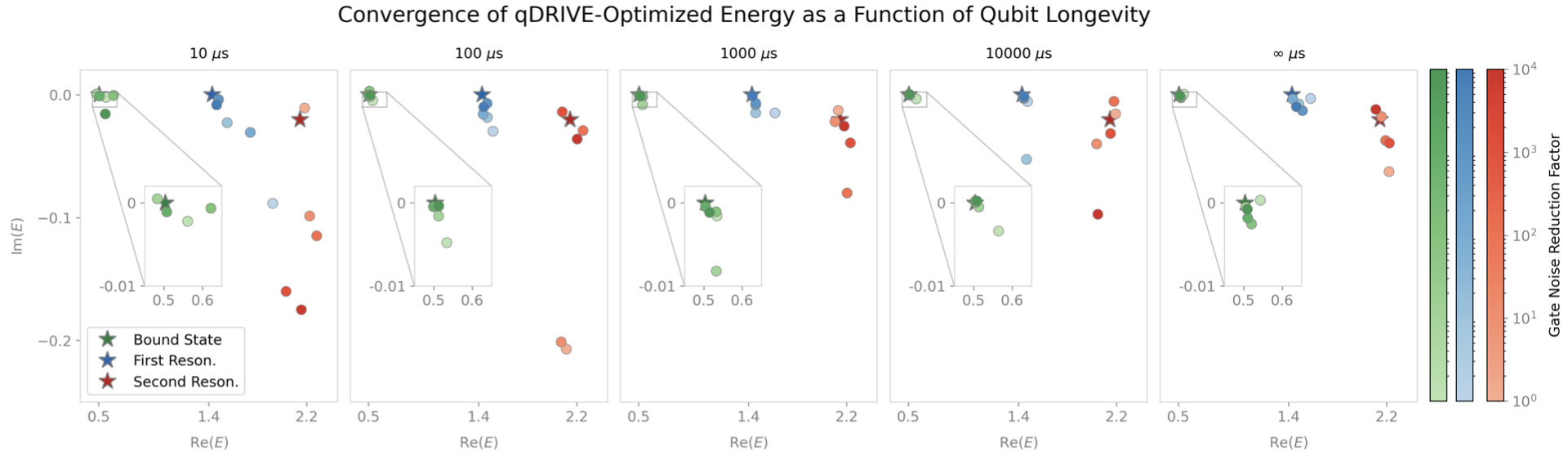}
\par\end{centering}
\caption{Approach of the bound-state (with magnification), first-resonance,
and second-resonance energy (circles; green, blue, and red, respectively)
to the corresponding \textcolor{black}{classical CAP method/}exact diagonalization energy (stars) with increased
qubit longevity. Darker colors indicate higher gate noise reduction
factor.\label{fig:QubitLongevity}}
\end{figure*}

\begin{figure*}
\begin{centering}
\includegraphics[width=1\textwidth]{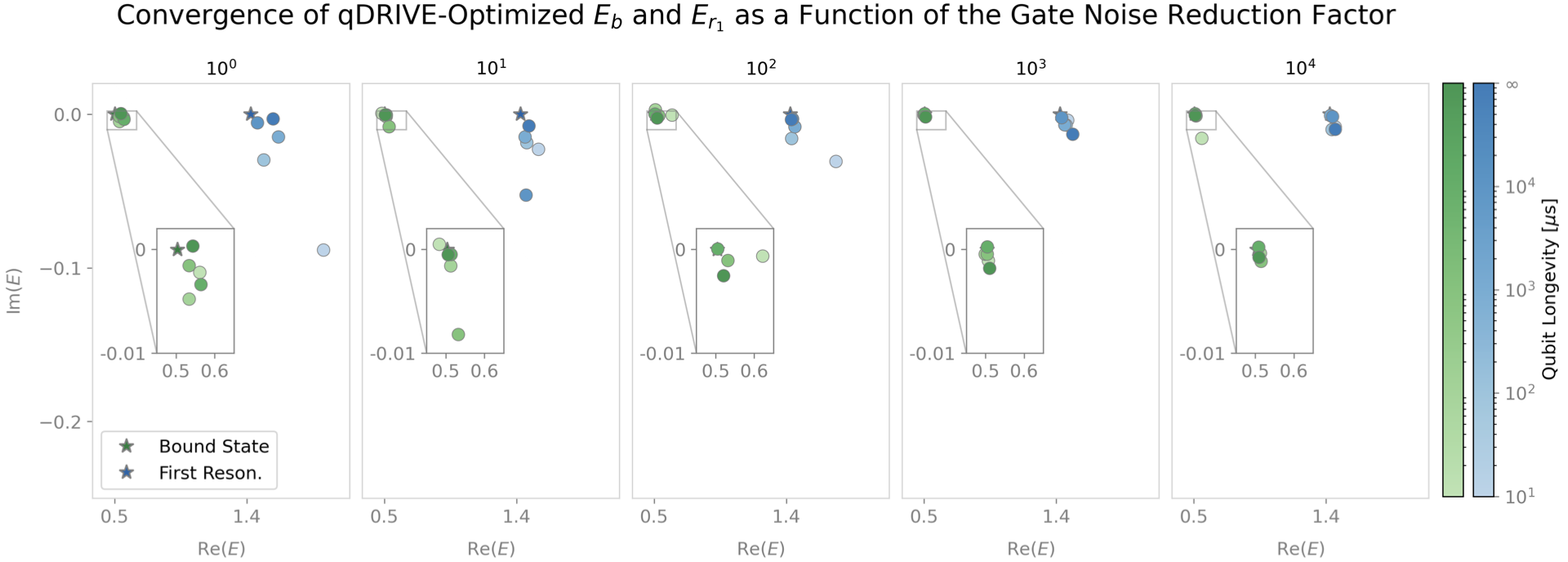}
\par\end{centering}
\caption{Progression of the energy of the bound state and first resonance (green
and blue circles, respectively) towards the corresponding \textcolor{black}{classical CAP method/}exact diagonalization
energy (green and blue stars, respectively) as the gate noise reduction
factor increases. Darker colors indicate longer qubit longevity.\label{fig:GateNoiseGroundFirst}}
\end{figure*}

\begin{figure*}
\begin{centering}
\includegraphics[width=1\textwidth]{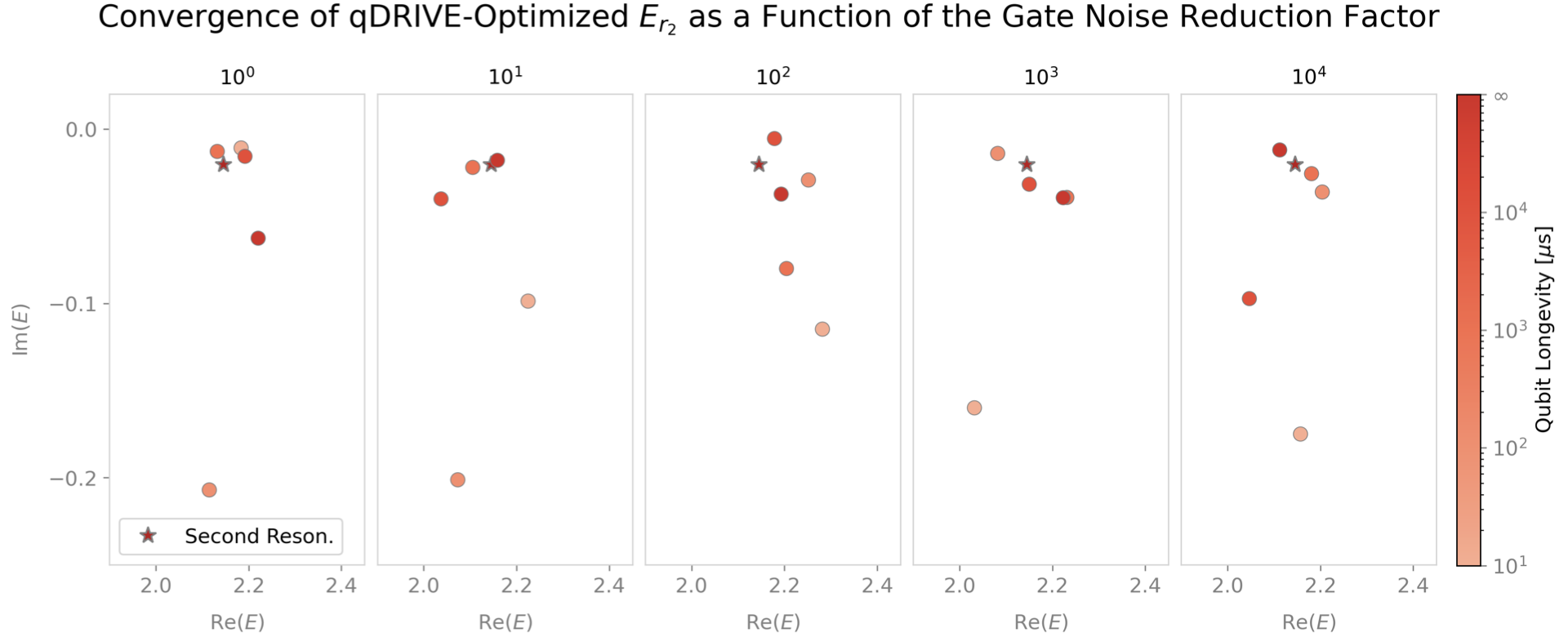}
\par\end{centering}
\caption{As in Fig.~\ref{fig:GateNoiseGroundFirst}, for the second resonance.\label{fig:GateNoiseSecond}}
\end{figure*}

Notably, convergence of the complex energies to their \textcolor{black}{classical CAP method}
values was evident even where the pseudovariance and fidelity errors
were far from their optimal zero values, as depicted in Figs.~\ref{fig:QubitLongevity},~\ref{fig:GateNoiseGroundFirst},
and~\ref{fig:GateNoiseSecond} for a three-qubit Ansatz. As shown
in Fig.~\ref{fig:QubitLongevity}, the real part of the qDRIVE-optimized
bound-state, first-resonance, and second-resonance energies closely
agreed with the \textcolor{black}{classical CAP method} value for all gate noise reduction
and qubit longevity factors considered, with imaginary parts with an
absolute error within $0.1$ regardless of gate noise for qubit
longevity factors of at least $1000\text{  \textmu s}$, on the order of
the millisecond times recently achieved on advanced 2D transmon qubits.\cite{bland2025millisecond}
The bound-state energy specifically was found to be nearly indistinguishable
from the \textcolor{black}{classical CAP method} result at the resolution shown without
magnification for qubit longevity factors of at least $10\text{  \textmu s}$,
namely, on the order of IBM Torino relaxation times.
As emphasized in Fig.~\ref{fig:GateNoiseGroundFirst}, the bound-state
and first-resonance energies' imaginary parts were accurate to within an absolute error of
$0.1$ for gate noise reduction factors of at least $10^{0}$
and qubit longevity factors of at least $10\text{  \textmu s}$, both of
which are accessible with current IBM Torino systems. Consistent with
trends in the pseudovariance and fidelity error, higher errors in
the real and imaginary parts of the energy were observed for the second
resonance associated with a higher-lying eigenstate of the Hermitian
Hamiltonian. As shown in Fig.~\ref{fig:GateNoiseSecond}, weaker
nonmonotonic convergence of the imaginary part of the qDRIVE-optimized
energy to \textcolor{black}{classical CAP method/}exact diagonalization results was observed for the second
resonance as the gate noise reduction factor increased, a correlation
between repeated deflation and error emergence that suggests possible
benefits of the future injection into qDRIVE of higher-accuracy VQD
variants (such as the tangent-vector VQE method\cite{wakaura2021tangent})
or variational algorithms for high-lying eigenstate identification
(such as adaptive VQE-X\cite{zhang2021adaptive}).

\section{Discussion and Conclusion}

The success of the proposed qDRIVE algorithm points to the potential
efficiency and efficacy of a hybrid quantum computing/high-throughput
computing (HTC) approach to chemistry. qDRIVE's ability to identify
molecular resonance energies and wavefunctions in a benchmark potential
on simulated noisy intermediate scale quantum (NISQ) computers
invites its implementation on today's quantum processors. Its projected
accuracy for more advanced quantum processors suggests the algorithm
may offer yet improved accuracy as quantum processors continue to
develop. And, more broadly, qDRIVE's success serves as a prototype
for a wider family of heterogeneous quantum computing algorithms. 

Making qDRIVE fully practical for systems beyond the small-scale systems
considered here will require addressing the issue of Hamiltonian mapping.
It is well-known that grid-based mappings, such as the method used
here to represent the benchmark potential, feature a storage cost
that grows exponentially with the grid size.\cite{mcardle2020quantum,kassal2011simulating}
Molecular electronic resonances\cite{Jagau2017ExtendingQC,jagau2022theory}
may be able to circumvent this problem given the existence of alternative,
scalable Hamiltonian mappings for electronic structure problems\cite{bauer2020quantum}
such as the renowned Jordan-Wigner\cite{jordan1928uber} and Bravyi-Kitaev\cite{bravyi2002fermionic}
fermion-qubit mappings. Additionally, where decompositions of the
Hamiltonian and its products result in a large number of expectation
values to be measured, a variety of methods currently under development
may reduce the number of expectation values that need to be computed,
and therefore reduce the overall cost of qDRIVE's implementation.
Classical shadow tomography,\cite{aaronson2018shadow,aaronson2019gentle,huang2020predicting}
for example, allows for the reconstruction of unmeasured expectation
values based on the ``shadow'' of a smaller subset of measured
values, and the method of commuting strings\cite{mcclean2016theory}
allows for large numbers of expectation values to be computed with
comparatively fewer quantum circuits.

An adaptive Ansatz approach offers a pathway to the extend qDRIVE
to larger-scale systems.  qDRIVE's observed ability to accurately
identify resonance energies and wavefunctions suggests the three-layer
efficient SU(2) Ansatz is sufficient for analysis of the benchmark
molecular system of interest; however, a key question is how to maintain
this expressibility and efficiency upon the addition of parameters
and entanglement between large numbers of qubits. One possibility
is ADAPT-VQE (Adaptive Derivative-Assembled Pseudo-Trotter Ansatz
Variational Quantum Eigensolver), which has been seen to be a highly
effective means to improve the scalability of VQE.\cite{grimsley2019adaptive}
By optimizing both the parameters and form of the Ansatz, a variant
of ADAPT-VQE has successfully enabled VQE on 127-qubit processors,
including a 100-qubit study of the lattice Schwinger model.\cite{farrell2024scalable}

\textcolor{black}{The contracted quantum eigensolver (CQE)\cite{smart2021quantum,wang2023electronic,wang2023quantum,warren2024exact} 
provides another pathway to expand qDRIVE to larger system sizes. 
As in ADAPT-VQE, CQE possesses the ability to modify an Ansatz until it suits the quantum system under study.
However, unlike ADAPT-VQE, CQE's two-body Ansatz construction method never calls for parameter reoptimization. 
This distinction, along with the structure of the contracted Schr{\"o}dinger equation on which CQE relies, allows CQE to readily produce analytic gradients 
of the energy with respect to the Ansatz parameters. 
Thus, whereas ADAPT-VQE typically relies on gradient-free classical optimization algorithms such as COBYLA, CQE can capitalize on efficient gradient-based classical optimization algorithms that offer important cost savings where quantum resources remain scarce.}

\textcolor{black}{In established classical resonance identification techniques, the computational cost generally scales 
directly with the size of the Hamiltonian matrix. For example, the matrix diagonalizations frequently used in classical CAP and complex scaling methods\cite{moiseyev1978resonance,moiseyev1998quantum} scale as $O\left(N^3\right)$. In contrast, the computational cost
of qDRIVE can scale independently of the Hamiltonian matrix size. qDRIVE's variational hybrid quantum-classical computing approach is expected to give qDRIVE a computational complexity akin to that of VQE,\cite{peruzzo2014variational}
namely, the cost of identifying of a single state in qDRIVE is expected to scale as $O\left(\left| C_\text{max} \right|^2 t p^{-2}\right)$, where $C_\text{max}$ is the largest-magnitude coefficient of the Pauli decomposition
Eq.~(\ref{eq:PauliDecomposition}), $t$ is the number of terms in said decomposition, and $p$ is the desired precision.\footnote{Note identification of additional states with qDRIVE is expected to indirectly depend on the matrix size due to the need for successive deflations. However, this dependence may be removed via substitution of the VQD routine with a suitable alternative for high-lying eigenstate identification. Likewise, although at its face the number of terms appears to scale with the matrix size as $t\propto N^2$, the number of terms is generically reducible to
the number of groups of commuting Pauli strings according to the method of commuting strings\cite{mcclean2016theory} 
and in many cases the number of terms is naturally significantly lower due to fortuitous construction of the Hamiltonian.}
The cost comparison therefore suggests the value of a search for possible advantage in the large system limit, 
with the caveat that the quantum utility of variational quantum algorithms remains a topic of continued heated debate.\cite{cerezo2021variational,Tilly2022Variational}}

qDRIVE's success invites the further development of integrated quantum
computing and HTC approaches. The interconnection of high-performance
computing (HPC) resources and quantum computers is an emerging hot
topic in quantum information science, and much focus has been placed
on distributed quantum computing systems based either on multiple
quantum processing units working in concert or heterogeneous computing
systems consisting of central processing units (CPUs), graphical processing
units (GPUs), and quantum processing units (QPUs).\cite{barral2025review,elsharkawy2025integration,rallis2025interfacing}
The accomplishments here of qDRIVE for molecular resonance identification,
paired with the accomplishments of HTC in classical scientific computing
for applications ranging from biochemical network\cite{kent2012condor}
to high-energy particle physics\cite{tsaregorodtsev2004dirac} simulations,
open the door to a powerful strategy of executing computational tasks
in parallel yet taking advantage of asynchronicity.

\section{Acknowledgements}

The authors thank Preetham Tikkireddi for insight into Qiskit implementation,
John Hawthorne for contributions to early versions of the code,
and the University of Wisconsin-Madison Center for High-Throughput
Computing and HTCondor Team for support with high-throughput computing
implementation. Support for this research was provided by the Office
of the Vice Chancellor for Research and Graduate Education at the
University of Wisconsin-Madison with funding from the Wisconsin Alumni
Research Foundation and with quantum computing resources from IBM
Quantum and the Chicago Quantum Exchange.

\bibliography{qdrive}

@article{wang2023quantum,
  title={Quantum simulation of bosons with the contracted quantum eigensolver},
  author={Wang, Yuchen and Sager-Smith, LeeAnn M and Mazziotti, David A},
  journal={New Journal of Physics},
  volume={25},
  number={10},
  pages={103005},
  year={2023},
  publisher={IOP Publishing}
}

@article{moiseyev1980association,
  title={Association of resonance states with the incomplete spectrum of finite complex-scaled Hamiltonian matrices},
  author={Moiseyev, Nimrod and Friedland, Shmuel},
  journal={Physical Review A},
  volume={22},
  number={2},
  pages={618},
  year={1980},
  publisher={APS}
}

@article{konotop2016nonlinear,
  title={Nonlinear waves in PT-symmetric systems},
  author={Konotop, Vladimir V and Yang, Jianke and Zezyulin, Dmitry A},
  journal={Reviews of Modern Physics},
  volume={88},
  number={3},
  pages={035002},
  year={2016},
  publisher={APS}
}

@article{ozdemir2019parity,
  title={Parity--time symmetry and exceptional points in photonics},
  author={{\"O}zdemir, {\c{S}}ahin Kaya and Rotter, Stefan and Nori, Franco and Yang, L},
  journal={Nature Materials},
  volume={18},
  number={8},
  pages={783--798},
  year={2019},
  publisher={Nature Publishing Group UK London}
}

@article{miri2019exceptional,
  title={Exceptional points in optics and photonics},
  author={Miri, Mohammad-Ali and Alu, Andrea},
  journal={Science},
  volume={363},
  number={6422},
  pages={eaar7709},
  year={2019},
  publisher={American Association for the Advancement of Science}
}

@article{havlivcek2019supervised,
  title={Supervised learning with quantum-enhanced feature spaces},
  author={Havl{\'\i}{\v{c}}ek, Vojt{\v{e}}ch and C{\'o}rcoles, Antonio D and Temme, Kristan and Harrow, Aram W and Kandala, Abhinav and Chow, Jerry M and Gambetta, Jay M},
  journal={Nature},
  volume={567},
  number={7747},
  pages={209--212},
  year={2019},
  publisher={Nature Publishing Group UK London}
}

@article{mansfield2025first,
  title={First Practical Experiences Integrating Quantum Computers with {HPC} Resources: A Case Study With a 20-qubit Superconducting Quantum Computer},
  author={Mansfield, Eric and Seegerer, Stefan and Vesanen, Panu and Echavarria, Jorge and Mete, Burak and Farooqi, Muhammad Nufail and Schulz, Laura},
  journal={arXiv preprint arXiv:2509.12949},
  year={2025}
}

@article{yarkoni2022quantum,
  title={Quantum annealing for industry applications: Introduction and review},
  author={Yarkoni, Sheir and Raponi, Elena and B{\"a}ck, Thomas and Schmitt, Sebastian},
  journal={Reports on Progress in Physics},
  volume={85},
  number={10},
  pages={104001},
  year={2022},
  publisher={IOP Publishing}
}

@article{wasielewski2020exploiting,
  title={Exploiting chemistry and molecular systems for quantum information science},
  author={Wasielewski, Michael R and Forbes, Malcolm DE and Frank, Natia L and Kowalski, Karol and Scholes, Gregory D and Yuen-Zhou, Joel and Baldo, Marc A and Freedman, Danna E and Goldsmith, Randall H and Goodson III, Theodore and Kirk, Martin L and McCusker, James K and Ogilvie, Jennifer P and Shultz, David A and Stoll, Stefan and Whaley, K Birgitta},
  journal={Nature Reviews Chemistry},
  volume={4},
  number={9},
  pages={490--504},
  year={2020},
  publisher={Nature Publishing Group UK London}
}

@article{wakaura2021tangent,
  title={Tangent vector variational quantum eigensolver: A robust variational quantum eigensolver against the inaccuracy of derivative},
  author={Wakaura, Hikaru and Suksmono, Andriyan Bayu},
  journal={arXiv preprint arXiv:2105.01141},
  year={2021}
}

@article{zhang2022variational,
  title={Variational quantum eigensolvers by variance minimization},
  author={Zhang, Dan-Bo and Chen, Bin-Lin and Yuan, Zhan-Hao and Yin, Tao},
  journal={Chinese Physics B},
  volume={31},
  number={12},
  pages={120301},
  year={2022},
  publisher={IOP Publishing}
}

@article{jordan1928uber,
       author = {{Jordan}, P. and {Wigner}, E.},
        title = "{{\"U}ber das Paulische {\"A}quivalenzverbot}",
      journal = {Zeitschrift fur Physik},
         year = 1928,
        month = sep,
       volume = {47},
       number = {9-10},
        pages = {631-651},
          doi = {10.1007/BF01331938},
       adsurl = {https://ui.adsabs.harvard.edu/abs/1928ZPhy...47..631J}
}

@article{liu2023probing,
  title={Probing many-body localization by excited-state variational quantum eigensolver},
  author={Liu, Shuo and Zhang, Shi-Xin and Hsieh, Chang-Yu and Zhang, Shengyu and Yao, Hong},
  journal={Physical Review B},
  volume={107},
  number={2},
  pages={024204},
  year={2023},
  publisher={APS}
}

@article{boyd2022training,
  title={Training variational quantum circuits with {CoVaR}: Covariance root finding with classical shadows},
  author={Boyd, Gregory and Koczor, B{\'a}lint},
  journal={Physical Review X},
  volume={12},
  number={4},
  pages={041022},
  year={2022},
  publisher={APS}
}

@article{hobday2022variance,
  title={Variance minimisation on a quantum computer for nuclear structure},
  author={Hobday, Isaac and Stevenson, Paul and Benstead, James},
  journal={arXiv preprint arXiv:2209.07820},
  year={2022}
}

@article{zhang2021adaptive,
  title={Adaptive variational quantum eigensolvers for highly excited states},
  author={Zhang, Feng and Gomes, Niladri and Yao, Yongxin and Orth, Peter P and Iadecola, Thomas},
  journal={Physical Review B},
  volume={104},
  number={7},
  pages={075159},
  year={2021},
  publisher={APS}
}

@article{bravyi2002fermionic,
  title={Fermionic quantum computation},
  author={Bravyi, Sergey B and Kitaev, Alexei Yu},
  journal={Annals of Physics},
  volume={298},
  number={1},
  pages={210--226},
  year={2002},
  publisher={Elsevier}
}

@inproceedings{aaronson2018shadow,
  title={Shadow tomography of quantum states},
  author={Aaronson, Scott},
  booktitle={Proceedings of the 50th annual ACM SIGACT symposium on theory of computing},
  pages={325--338},
  year={2018}
}

@article{huang2020predicting,
  title={Predicting many properties of a quantum system from very few measurements},
  author={Huang, Hsin-Yuan and Kueng, Richard and Preskill, John},
  journal={Nature Physics},
  volume={16},
  number={10},
  pages={1050--1057},
  year={2020},
  publisher={Nature Publishing Group UK London}
}

@article{grimsley2025challenging,
  title={Challenging excited states from adaptive quantum eigensolvers: subspace expansions vs. state-averaged strategies},
  author={Grimsley, Harper R and Evangelista, Francesco A},
  journal={Quantum Science and Technology},
  volume={10},
  number={2},
  pages={025003},
  year={2025},
  publisher={IOP Publishing}
}

@article{farrell2024scalable,
  title={Scalable circuits for preparing ground states on digital quantum computers: The {Schwinger} model vacuum on 100 qubits},
  author={Farrell, Roland C and Illa, Marc and Ciavarella, Anthony N and Savage, Martin J},
  journal={PRX Quantum},
  volume={5},
  number={2},
  pages={020315},
  year={2024},
  publisher={APS}
}

@article{grimsley2019adaptive,
  title={An adaptive variational algorithm for exact molecular simulations on a quantum computer},
  author={Grimsley, Harper R and Economou, Sophia E and Barnes, Edwin and Mayhall, Nicholas J},
  journal={Nature Communications},
  volume={10},
  number={1},
  pages={3007},
  year={2019},
  publisher={Nature Publishing Group UK London}
}

@inproceedings{aaronson2019gentle,
  title={Gentle measurement of quantum states and differential privacy},
  author={Aaronson, Scott and Rothblum, Guy N},
  booktitle={Proceedings of the 51st Annual ACM SIGACT Symposium on Theory of Computing},
  pages={322--333},
  year={2019}
}

@article{mcclean2016theory,
  title={The theory of variational hybrid quantum-classical algorithms},
  author={McClean, Jarrod R and Romero, Jonathan and Babbush, Ryan and Aspuru-Guzik, Al{\'a}n},
  journal={New Journal of Physics},
  volume={18},
  number={2},
  pages={023023},
  year={2016},
  publisher={IOP Publishing}
}

@article{jagau2022theory,
  title={Theory of electronic resonances: fundamental aspects and recent advances},
  author={Jagau, Thomas-C},
  journal={Chemical Communications},
  volume={58},
  number={34},
  pages={5205--5224},
  year={2022},
  publisher={The Royal Society of Chemistry}
}

@article{kassal2011simulating,
  title={Simulating chemistry using quantum computers},
  author={Kassal, Ivan and Whitfield, James D and Perdomo-Ortiz, Alejandro and Yung, Man-Hong and Aspuru-Guzik, Al{\'a}n},
  journal={Annual Review of Physical Chemistry},
  volume={62},
  number={1},
  pages={185--207},
  year={2011},
  publisher={Annual Reviews}
}

@article{bland2025millisecond,
  title={Millisecond lifetimes and coherence times in {2D} transmon qubits},
  author={Matthew P Bland and Faranak Bahrami and Jeronimo GC Martinez and Paal H. Prestegaard and Basil M. Smitham and Atharv Joshi and Elizabeth Hedrick and Shashwat Kumar and Ambrose Yang and Alexander C Pakpour-Tabrizi and Apoorv Jindal and Ray D Chang and Guangming Cheng and Nan Yao and Robert J Cava and Nathalie P de Leon and Andrew A Houck},
  journal={Nature},
  volume={},
  number={},
  pages={},
  year={2025},
  publisher={Nature Publishing Group UK London}
}

@INPROCEEDINGS{litzkow1987condor,
  author={Litzkow, M.J. and Livny, M. and Mutka, M.W.},
  booktitle={The 8th International Conference on Distributed Computing Systems}, 
  title={Condor-a hunter of idle workstations}, 
  year={1988},
  volume={},
  number={},
  pages={104-111},
  doi={10.1109/DCS.1988.12507}
}

@inproceedings{fajardo2015much,
  title={How much higher can HTCondor fly?},
  author={Fajardo, EM and Dost, JM and Holzman, B and Tannenbaum, T and Letts, J and Tiradani, A and Bockelman, B and Frey, J and Mason, D},
  booktitle={Journal of Physics: Conference Series},
  volume={664},
  number={6},
  pages={062014},
  year={2015},
  organization={IOP Publishing}
}

@inproceedings{litzkow1987remote,
  title={Remote Unix: Turning idle workstations into cycle servers},
  author={Litzkow, Michael J},
  booktitle={Proceedings of the Summer USENIX Conference},
  pages={381--384},
  year={1987}
}

@misc{osgconnectgithub,
  author       = {Open Science Grid},
  title        = {{Intermediate DAGMan}},
  howpublished = {\url{https://github.com/OSGConnect/tutorial-dagman-intermediate}},
  year         = {2024}
}

@article{isakov2021simulations,
  title={Simulations of quantum circuits with approximate noise using qsim and {Cirq}},
  author={Isakov, Sergei V and Kafri, Dvir and Martin, Orion and Heidweiller, Catherine Vollgraff and Mruczkiewicz, Wojciech and Harrigan, Matthew P and Rubin, Nicholas C and Thomson, Ross and Broughton, Michael and Kissell, Kevin and Peters, Evan and Gustafson, Erik and Li, Andy C Y and Lamm, Henry and Perdue, Gabriel and Ho, Alan K and Strain, Doug and Boixo, Sergio},
  journal={arXiv preprint arXiv:2111.02396},
  year={2021}
}

@misc{qdrivegithub,
  author       = {Jingcheng Dai and Atharva Vidwans and Eric H Wan and Alexander X Miller and John M Hawthorne and Micheline B Soley},
  title        = {{qDRIVE}},
  howpublished = {\url{https://github.com/ericwan808/qDRIVE}},
  year         = {2026}
}

@article{javadi2024quantum,
  title={Quantum computing with {Qiskit}},
  author={Javadi-Abhari, Ali and Treinish, Matthew and Krsulich, Kevin and Wood, Christopher J and Lishman, Jake and Gacon, Julien and Martiel, Simon and Nation, Paul D and Bishop, Lev S and Cross, Andrew W and Johnson, Blake R and Gambetta, Jay M},
  journal={arXiv preprint arXiv:2405.08810},
  year={2024}
}

@article{maciejewski2020mitigation,
  title={Mitigation of readout noise in near-term quantum devices by classical post-processing based on detector tomography},
  author={Maciejewski, Filip B and Zimbor{\'a}s, Zolt{\'a}n and Oszmaniec, Micha{\l}},
  journal={Quantum},
  volume={4},
  pages={257},
  year={2020},
  publisher={Verein zur F{\"o}rderung des Open Access Publizierens in den Quantenwissenschaften}
}

@inproceedings{powell1994direct,
  title={A direct search optimization method that models the objective and constraint functions by linear interpolation},
  author={Powell, Michael JD},
  booktitle={Advances in optimization and numerical analysis},
  pages={51--67},
  year={1994},
  publisher={Springer}
}

@article{herzenberg1963resonant,
  title={Resonant electron scattering by atoms},
  author={Herzenberg, A and Mandl, F},
  journal={Proceedings of the Royal Society A},
  volume={274},
  number={1357},
  pages={253--266},
  year={1963},
  publisher={The Royal Society London}
}

@article{buhrman2001quantum,
  title={Quantum fingerprinting},
  author={Buhrman, Harry and Cleve, Richard and Watrous, John and De Wolf, Ronald},
  journal={Physical Review Letters},
  volume={87},
  number={16},
  pages={167902},
  year={2001},
  publisher={APS}
}

@article{gottesman2001quantum,
  title={Quantum digital signatures},
  author={Gottesman, Daniel and Chuang, Isaac},
  journal={arXiv preprint quant-ph/0105032},
  year={2001}
}

@article{cincio2018learning,
  title={Learning the quantum algorithm for state overlap},
  author={Cincio, Lukasz and Suba{\c{s}}{\i}, Yi{\u{g}}it and Sornborger, Andrew T and Coles, Patrick J},
  journal={New Journal of Physics},
  volume={20},
  number={11},
  pages={113022},
  year={2018},
  publisher={IOP Publishing}
}

@article{garcia2013swap,
  title={{SWAP} test and {Hong-Ou-Mandel} effect are equivalent},
  author={Garcia-Escartin, Juan Carlos and Chamorro-Posada, Pedro},
  journal={Physical Review A},
  volume={87},
  number={5},
  pages={052330},
  year={2013},
  publisher={APS}
}

@article{cerezo2021variational,
  title={Variational quantum algorithms},
  author={Cerezo, Marco and Arrasmith, Andrew and Babbush, Ryan and Benjamin, Simon C and Endo, Suguru and Fujii, Keisuke and McClean, Jarrod R and Mitarai, Kosuke and Yuan, Xiao and Cincio, Lukasz and Coles, Patrick J},
  journal={Nature Reviews Physics},
  volume={3},
  number={9},
  pages={625--644},
  year={2021},
  publisher={Nature Publishing Group UK London}
}

@article{dewes2012characterization,
  title={Characterization of a two-transmon processor with individual single-shot qubit readout},
  author={Dewes, Andreas and Ong, Florian R and Schmitt, Vivien and Lauro, R and Boulant, N and Bertet, P and Vion, D and Esteve, D},
  journal={Physical Review Letters},
  volume={108},
  number={5},
  pages={057002},
  year={2012},
  publisher={APS}
}

@article{van2022model,
  title={Model-free readout-error mitigation for quantum expectation values},
  author={Van Den Berg, Ewout and Minev, Zlatko K and Temme, Kristan},
  journal={Physical Review A},
  volume={105},
  number={3},
  pages={032620},
  year={2022},
  publisher={APS}
}

@article{karalekas2020quantum,
  title={A quantum-classical cloud platform optimized for variational hybrid algorithms},
  author={Karalekas, Peter J and Tezak, Nikolas A and Peterson, Eric C and Ryan, Colm A and Da Silva, Marcus P and Smith, Robert S},
  journal={Quantum Science and Technology},
  volume={5},
  number={2},
  pages={024003},
  year={2020},
  publisher={IOP Publishing}
}

@article{kandala2019error,
  title={Error mitigation extends the computational reach of a noisy quantum processor},
  author={Kandala, Abhinav and Temme, Kristan and C{\'o}rcoles, Antonio D and Mezzacapo, Antonio and Chow, Jerry M and Gambetta, Jay M},
  journal={Nature},
  volume={567},
  number={7749},
  pages={491--495},
  year={2019},
  publisher={Nature Publishing Group UK London}
}

@article{miller2025universal,
  title={Universal Quantum Error Mitigation via Random Inverse Depolarizing Approximation},
  author={Miller, Alexander X and Soley, Micheline B},
  journal={arXiv preprint arXiv:2508.17513},
  year={2025}
}

@article{endo2018practical,
  title={Practical quantum error mitigation for near-future applications},
  author={Endo, Suguru and Benjamin, Simon C and Li, Ying},
  journal={Physical Review X},
  volume={8},
  number={3},
  pages={031027},
  year={2018},
  publisher={APS}
}

@article{cai2023quantum,
  title={Quantum error mitigation},
  author={Cai, Zhenyu and Babbush, Ryan and Benjamin, Simon C and Endo, Suguru and Huggins, William J and Li, Ying and McClean, Jarrod R and O'Brien, Thomas E},
  journal={Reviews of Modern Physics},
  volume={95},
  number={4},
  pages={045005},
  year={2023},
  publisher={APS}
}

@article{livshits1957application,
  title={The application of non-self-adjoint operators to scattering theory},
  author={Livshits, MS},
  journal={Soviet Physics, Journal of Experimental and Theoretical Physics},
  volume={4},
  pages={91--98},
  year={1957}
}

@article{feshbach1958unified,
  title={Unified theory of nuclear reactions},
  author={Feshbach, Herman},
  journal={Annals of Physics},
  volume={5},
  number={4},
  pages={357--390},
  year={1958},
  publisher={Elsevier}
}

@article{zhang2025iterative,
  title={Iterative {Harrow-Hassidim-Lloyd} quantum algorithm for solving resonances with eigenvector continuation},
  author={Zhang, Hantao and Bai, Dong and Ren, Zhongzhou},
  journal={arXiv preprint arXiv:2506.20929},
  year={2025}
}

@article{feshbach1962unified,
  title={A unified theory of nuclear reactions. {II}},
  author={Feshbach, Herman},
  journal={Annals of Physics},
  volume={19},
  number={2},
  pages={287--313},
  year={1962},
  publisher={Elsevier}
}

@article{kim2023two,
  title={Two algorithms for excited-state quantum solvers: Theory and application to {EOM-UCCSD}},
  author={Kim, Yongbin and Krylov, Anna I},
  journal={Journal of Physical Chemistry A},
  volume={127},
  number={31},
  pages={6552--6566},
  year={2023},
  publisher={ACS Publications}
}

@article{wang2023electronic,
  title={Electronic excited states from a variance-based contracted quantum eigensolver},
  author={Wang, Yuchen and Mazziotti, David A},
  journal={Physical Review A},
  volume={108},
  number={2},
  pages={022814},
  year={2023},
  publisher={APS}
}

@article{nakanishi2019subspace,
  title={Subspace-search variational quantum eigensolver for excited states},
  author={Nakanishi, Ken M and Mitarai, Kosuke and Fujii, Keisuke},
  journal={Physical Review Research},
  volume={1},
  number={3},
  pages={033062},
  year={2019},
  publisher={APS}
}

@article{cianci2024subspace,
  title={Subspace-Search Quantum Imaginary Time Evolution for Excited State Computations},
  author={Cianci, Cameron and Santos, Lea F and Batista, Victor S},
  journal={Journal of Chemical Theory and Computation},
  volume={20},
  number={20},
  pages={8940--8947},
  year={2024},
  publisher={ACS Publications}
}

@article{ollitrault2020quantum,
  title={Quantum equation of motion for computing molecular excitation energies on a noisy quantum processor},
  author={Ollitrault, Pauline J and Kandala, Abhinav and Chen, Chun-Fu and Barkoutsos, Panagiotis Kl and Mezzacapo, Antonio and Pistoia, Marco and Sheldon, Sarah and Woerner, Stefan and Gambetta, Jay M and Tavernelli, Ivano},
  journal={Physical Review Research},
  volume={2},
  number={4},
  pages={043140},
  year={2020},
  publisher={APS}
}

@article{kuroiwa2021penalty,
  title={Penalty methods for a variational quantum eigensolver},
  author={Kuroiwa, Kohdai and Nakagawa, Yuya O},
  journal={Physical Review Research},
  volume={3},
  number={1},
  pages={013197},
  year={2021},
  publisher={APS}
}

@article{colless2018computation,
  title={Computation of molecular spectra on a quantum processor with an error-resilient algorithm},
  author={Colless, James I and Ramasesh, Vinay V and Dahlen, Dar and Blok, Machiel S and Kimchi-Schwartz, Mollie E and McClean, Jarrod R and Carter, Jonathan and de Jong, Wibe A and Siddiqi, Irfan},
  journal={Physical Review X},
  volume={8},
  number={1},
  pages={011021},
  year={2018},
  publisher={APS}
}

@article{santagati2018witnessing,
  title={Witnessing eigenstates for quantum simulation of Hamiltonian spectra},
  author={Santagati, Raffaele and Wang, Jianwei and Gentile, Antonio A and Paesani, Stefano and Wiebe, Nathan and McClean, Jarrod R and Morley-Short, Sam and Shadbolt, Peter J and Bonneau, Damien and Silverstone, Joshua W and Tew, David P. and Zhou, Xiaoqi and O'Brien, Jeremy L and Thompson, Mark G},
  journal={Science Advances},
  volume={4},
  number={1},
  pages={eaap9646},
  year={2018},
  publisher={American Association for the Advancement of Science}
}

@article{motta2020determining,
  title={Determining eigenstates and thermal states on a quantum computer using quantum imaginary time evolution},
  author={Motta, Mario and Sun, Chong and Tan, Adrian TK and O'Rourke, Matthew J and Ye, Erika and Minnich, Austin J and Brandao, Fernando GSL and Chan, Garnet Kin-Lic},
  journal={Nature Physics},
  volume={16},
  number={2},
  pages={205--210},
  year={2020},
  publisher={Nature Publishing Group UK London}
}

@article{kyaw2023boosting,
  title={Boosting quantum amplitude exponentially in variational quantum algorithms},
  author={Kyaw, Thi Ha and Soley, Micheline B and Allen, Brandon and Bergold, Paul and Sun, Chong and Batista, Victor S and Aspuru-Guzik, Al{\'a}n},
  journal={Quantum Science and Technology},
  volume={9},
  number={1},
  pages={01LT01},
  year={2023},
  publisher={IOP Publishing}
}

@article{hejazi2024adiabatic,
  title={Adiabatic quantum imaginary time evolution},
  author={Hejazi, Kasra and Motta, Mario and Chan, Garnet Kin-Lic},
  journal={Physical Review Research},
  volume={6},
  number={3},
  pages={033084},
  year={2024},
  publisher={APS}
}

@article{abrams1999quantum,
  title={Quantum algorithm providing exponential speed increase for finding eigenvalues and eigenvectors},
  author={Abrams, Daniel S and Lloyd, Seth},
  journal={Physical Review Letters},
  volume={83},
  number={24},
  pages={5162},
  year={1999},
  publisher={APS}
}

@article{bauer2020quantum,
  title={Quantum algorithms for quantum chemistry and quantum materials science},
  author={Bauer, Bela and Bravyi, Sergey and Motta, Mario and Chan, Garnet Kin-Lic},
  journal={Chemical Reviews},
  volume={120},
  number={22},
  pages={12685--12717},
  year={2020},
  publisher={ACS Publications}
}

@article{albash2018adiabatic,
  title={Adiabatic quantum computation},
  author={Albash, Tameem and Lidar, Daniel A},
  journal={Reviews of Modern Physics},
  volume={90},
  number={1},
  pages={015002},
  year={2018},
  publisher={APS}
}

@article{farhi2000quantum,
  title={Quantum computation by adiabatic evolution},
  author={Farhi, Edward and Goldstone, Jeffrey and Gutmann, Sam and Sipser, Michael},
  journal={arXiv preprint quant-ph/0001106},
  year={2000}
}

@article{peruzzo2014variational,
  title={A variational eigenvalue solver on a photonic quantum processor},
  author={Peruzzo, Alberto and McClean, Jarrod and Shadbolt, Peter and Yung, Man-Hong and Zhou, Xiao-Qi and Love, Peter J and Aspuru-Guzik, Al{\'a}n and O'Brien, Jeremy L},
  journal={Nature Communications},
  volume={5},
  number={1},
  pages={4213},
  year={2014},
  publisher={Nature Publishing Group UK London}
}

@article{smart2021quantum,
  title={Quantum solver of contracted eigenvalue equations for scalable molecular simulations on quantum computing devices},
  author={Smart, Scott E and Mazziotti, David A},
  journal={Physical Review Letters},
  volume={126},
  number={7},
  pages={070504},
  year={2021},
  publisher={APS}
}

@article{warren2024exact,
  title={Exact ansatz of fermion-boson systems for a quantum device},
  author={Warren, Samuel and Wang, Yuchen and Benavides-Riveros, Carlos L and Mazziotti, David A},
  journal={Physical Review Letters},
  volume={133},
  number={8},
  pages={080202},
  year={2024},
  publisher={APS}
}

@article{delgado2025quantum,
  title={Quantum algorithms and applications for open quantum systems},
  author={Delgado-Granados, Luis H and Krogmeier, Timothy J and Sager-Smith, LeeAnn M and Avdic, Irma and Hu, Zixuan and Sajjan, Manas and Abbasi, Maryam and Smart, Scott E and Narang, Prineha and Kais, Sabre and Schlimgen, Anthony W and Head-Marsden, Kade and Mazziotti, David A},
  journal={Chemical Reviews},
  volume={125},
  number={4},
  pages={1823--1839},
  year={2025},
  publisher={ACS Publications}
}

@article{aguilar1971class,
  title={A class of analytic perturbations for one-body {Schr{\"o}dinger} Hamiltonians},
  author={Aguilar, Jacques and Combes, Jean-Michel},
  journal={Communications in Mathematical Physics},
  volume={22},
  number={4},
  pages={269--279},
  year={1971},
  publisher={Springer}
}

@article{kukulin1977description,
  title={Description of few-body systems via analytical continuation in coupling constant},
  author={Kukulin, VI and Krasnopol'sky, VM},
  journal={Journal of Physics A: Mathematical and General},
  volume={10},
  number={2},
  pages={L33},
  year={1977},
  publisher={IOP Publishing}
}

@article{eliezer1967resonant,
  title={Resonant states of {$\text{H}_2^-$}},
  author={Eliezer, I and Taylor, HS and Williams, James K},
  journal={Journal of Chemical Physics},
  volume={47},
  number={6},
  pages={2165--2177},
  year={1967},
  publisher={AIP Publishing}
}

@article{taylor1966qualitative,
  title={Qualitative aspects of resonances in electron--atom and electron--molecule scattering, excitation, and reactions},
  author={Taylor, Howard S and Nazaroff, George V and Golebiewski, A},
  journal={Journal of Chemical Physics},
  volume={45},
  number={8},
  pages={2872--2888},
  year={1966},
  publisher={American Institute of Physics}
}

@inproceedings{taylor1970models,
	author = {Taylor, Howard S.},
	publisher = {John Wiley and Sons, Ltd},
	isbn = {9780470143650},
	title = {Models, Interpretations, and Calculations Concerning Resonant Electron Scattering Processes in Atoms and Molecules},
	booktitle = {Advances in Chemical Physics},
	chapter = {},
	pages = {91-147},
	doi = {https://doi.org/10.1002/9780470143650.ch3},
	url = {https://onlinelibrary.wiley.com/doi/abs/10.1002/9780470143650.ch3},
	eprint = {https://onlinelibrary.wiley.com/doi/pdf/10.1002/9780470143650.ch3},
	year = {1970}
}

@article{descouvemont2024resonances,
  title={Resonances in the R-matrix method},
  author={Descouvemont, Pierre and Dohet-Eraly, J{\'e}r{\'e}my},
  journal={Few-Body Systems},
  volume={65},
  number={1},
  pages={9},
  year={2024},
  publisher={Springer}
}

@article{kukulin1979method,
  title={Method of analytic continuation in the coupling constant in the theory of systems of several particles. {Resonance} state as analytic continuation of a bound state},
  author={Kukulin, VI and Krasnopol'skii, VM and Miselkhi, M},
  journal={Soviet Journal of Nuclear Physics (English Translation)},
  volume={29},
  number={3},
  year={1979},
  publisher={Institute of Nuclear Physics, Moscow State University}
}

@article{balslev1971spectral,
  title={Spectral properties of many-body {Schr{\"o}dinger} operators with dilatation-analytic interactions},
  author={Balslev, Erik and Combes, Jean-Michel},
  journal={Communications in Mathematical Physics},
  volume={22},
  number={4},
  pages={280--294},
  year={1971},
  publisher={Springer}
}

@article{mcardle2020quantum,
  title={Quantum computational chemistry},
  author={McArdle, Sam and Endo, Suguru and Aspuru-Guzik, Al{\'a}n and Benjamin, Simon C and Yuan, Xiao},
  journal={Reviews of Modern Physics},
  volume={92},
  number={1},
  pages={015003},
  year={2020},
  publisher={APS}
}

@article{cao2019quantum,
  title={Quantum chemistry in the age of quantum computing},
  author={Cao, Yudong and Romero, Jonathan and Olson, Jonathan P and Degroote, Matthias and Johnson, Peter D and Kieferov{\'a}, M{\'a}ria and Kivlichan, Ian D and Menke, Tim and Peropadre, Borja and Sawaya, Nicolas PD and Sim, Sukin and Veis, Libor and Aspuru-Guzik, Al{\'a}n},
  journal={Chemical Reviews},
  volume={119},
  number={19},
  pages={10856--10915},
  year={2019},
  publisher={ACS Publications}
}

@article{neuhasuer1989time,
  title={The time-dependent Schr{\"o}dinger equation: Application of absorbing boundary conditions},
  author={Neuhasuer, Daniel and Baer, Michael},
  journal={Journal of Chemical Physics},
  volume={90},
  number={8},
  pages={4351--4355},
  year={1989},
  publisher={AIP Publishing}
}

@article{leforestier1983optical,
  title={Optical potential for laser induced dissociation},
  author={Leforestier, Claude and Wyatt, Robert E},
  journal={Journal of Chemical Physics},
  volume={78},
  number={5},
  pages={2334--2344},
  year={1983},
  publisher={American Institute of Physics}
}

@article{leforestier1985role,
  title={Role of Feshbach resonances in the infrared multiphoton dissociation of small molecules},
  author={Leforestier, Claude and Wyatt, Robert E},
  journal={Journal of Chemical Physics},
  volume={82},
  number={2},
  pages={752--757},
  year={1985},
  publisher={American Institute of Physics}
}

@article{kosloff1986absorbing,
  title={Absorbing boundaries for wave propagation problems},
  author={Kosloff, R and Kosloff, D},
  journal={Journal of Computational Physics},
  volume={63},
  number={2},
  pages={363--376},
  year={1986},
  publisher={Elsevier}
}

@article{jolicard1985optical,
  title={Optical potential stabilisation method for predicting resonance levels},
  author={Jolicard, Georges and Austin, Elizabeth J},
  journal={Chemical Physics Letters},
  volume={121},
  number={1-2},
  pages={106--110},
  year={1985},
  publisher={Elsevier}
}

@article{kosloff1984dynamical,
  title={Dynamical atom/surface effects: Quantum mechanical scattering and desorption},
  author={Kosloff, Ronnie and Cerjan, Charles},
  journal={Journal of Chemical Physics},
  volume={81},
  number={8},
  pages={3722--3729},
  year={1984},
  publisher={American Institute of Physics}
}

@article{teplukhin2020solving,
  title={Solving complex eigenvalue problems on a quantum annealer with applications to quantum scattering resonances},
  author={Teplukhin, Alexander and Kendrick, Brian K and Babikov, Dmitri},
  journal={Physical Chemistry Chemical Physics},
  volume={22},
  number={45},
  pages={26136--26144},
  year={2020},
  publisher={Royal Society of Chemistry}
}

@inproceedings{giurgica2020digital,
  title={Digital zero noise extrapolation for quantum error mitigation},
  author={Giurgica-Tiron, Tudor and Hindy, Yousef and LaRose, Ryan and Mari, Andrea and Zeng, William J},
  booktitle={2020 IEEE International Conference on Quantum Computing and Engineering},
  pages={306--316},
  year={2020},
  organization={IEEE}
}

@article{temme2017error,
  title={Error mitigation for short-depth quantum circuits},
  author={Temme, Kristan and Bravyi, Sergey and Gambetta, Jay M},
  journal={Physical Review Letters},
  volume={119},
  number={18},
  pages={180509},
  year={2017},
  publisher={APS}
}

@article{geller2020rigorous,
  title={Rigorous measurement error correction},
  author={Geller, Michael R},
  journal={Quantum Science and Technology},
  volume={5},
  number={3},
  pages={03LT01},
  year={2020},
  publisher={IOP Publishing}
}

@article{hancock2025quantum,
  title={Quantum Phase Transition of Non-Hermitian Systems using Variational Quantum Techniques},
  author={Hancock, James and Craven, Matthew J and McNeile, Craig and Vadacchino, Davide},
  journal={arXiv preprint arXiv:2501.17003},
  year={2025}
}

@article{singh2024quantum,
    author = {Singh, Ashutosh and Maheshwar, R. and Siwach, P. and Singh, N. and Arumugam, P.},
    title = {Quantum Simulation of Complex-Scaled {Hamiltonian}},
    journal = {Department of Atomic Energy Symposium on Nuclear Physics},
    volume = {67},
    pages = {209--210},
    year = {2024}
}

@article{Jagau2017ExtendingQC,
  title={Extending Quantum Chemistry of Bound States to Electronic Resonances.},
  author={Thomas-C. Jagau and Ksenia B. Bravaya and Anna I. Krylov},
  journal={Annual Review of Physical Chemistry},
  year={2017},
  volume={68},
  pages={525-553},
  url={https://api.semanticscholar.org/CorpusID:12224757}
}

@article{barral2025review,
  title={Review of distributed quantum computing: from single QPU to high performance quantum computing},
  author={Barral, David and Cardama, F Javier and Diaz-Camacho, Guillermo and Fa{\'\i}lde, Daniel and Llovo, Iago F and Mussa-Juane, Mariamo and V{\'a}zquez-P{\'e}rez, Jorge and Villasuso, Juan and Pi{\~n}eiro, C{\'e}sar and Costas, Natalia and Pichel, Juan C and Pena, Tom{\'a}s F and G{\'o}mez, Andr{\'e}s},
  journal={Computer Science Review},
  volume={57},
  pages={100747},
  year={2025},
  publisher={Elsevier}
}

@article{kent2012condor,
  title={{Condor-COPASI}: high-throughput computing for biochemical networks},
  author={Kent, Edward and Hoops, Stefan and Mendes, Pedro},
  journal={BMC Systems Biology},
  volume={6},
  number={1},
  pages={91},
  year={2012},
  publisher={Springer}
}

@inproceedings{tsaregorodtsev2004dirac,
  title={DIRAC: A scalable lightweight architecture for high throughput computing},
  author={Tsaregorodtsev, Andrei and Garonne, Vincent and Stokes-Rees, Ian},
  booktitle={Fifth IEEE/ACM International Workshop on Grid Computing},
  pages={19--25},
  year={2004},
  organization={IEEE}
}

@article{altunay2011science,
  title={A science driven production cyberinfrastructure--the open science grid},
  author={Altunay, Mine and Avery, Paul and Blackburn, Kent and Bockelman, Brian and Ernst, Michael and Fraser, Dan and Quick, Robert and Gardner, Robert and Goasguen, Sebastien and Levshina, Tanya and Livny, Miron and McGee, John and Olson, Doug and Pordes, Ruth and Potekhin, Maxim and Rana, Abhishek and Roy, Alain and Sehgal, Chander and Sfiligoi, Igor and Wuerthwein, Frank},
  journal={Journal of Grid Computing},
  volume={9},
  number={2},
  pages={201--218},
  year={2011},
  publisher={Springer}
}

@inproceedings{pordes2007open,
  title={The open science grid},
  author={Pordes, Ruth and Petravick, Don and Kramer, Bill and Olson, Doug and Livny, Miron and Roy, Alain and Avery, Paul and Blackburn, Kent and Wenaus, Torre and W{\"u}rthwein, Frank and Foster, Ian and Gardner, Rob and Wilde, Mike and Blatecky, Alan and McGee, John and Quick, Rob},
  booktitle={Journal of Physics: Conference Series},
  volume={78},
  number={1},
  year={2007}
}

@article{rallis2025interfacing,
  title={Interfacing Quantum Computing Systems with High-Performance Computing Systems: An Overview},
  author={Rallis, Konstantinos and Liliopoulos, Ioannis and Varsamis, Georgios D and Tsipas, Evangelos and Karafyllidis, Ioannis G and Sirakoulis, Georgios Ch and Dimitrakis, Panagiotis},
  journal={arXiv preprint arXiv:2509.06205},
  year={2025}
}

@article{elsharkawy2025integration,
  title={Integration of quantum accelerators with high performance computing--a review of quantum programming tools},
  author={Elsharkawy, Amr and To, Xiao-Ting Michelle and Seitz, Philipp and Chen, Yanbin and Stade, Yannick and Geiger, Manuel and Huang, Qunsheng and Guo, Xiaorang and Ansari, Muhammad Arslan and Mendl, Christian B and Kranzlm{\"u}ller, Dieter and Schulz, Martin},
  journal={ACM Transactions on Quantum Computing},
  volume={6},
  number={3},
  pages={1--46},
  year={2025},
  publisher={ACM New York, NY}
}

@book{Moiseyev2011NonHermitian,
place={Cambridge},
title={Non-Hermitian Quantum Mechanics},
publisher={Cambridge University Press},
author={Moiseyev Nimrod},
year={2011}
}

@article{moiseyev1978resonance,
author = {Moiseyev, Nimrod and Certain, Phillip and Weinhold, Frank},
year = {1978},
month = {12},
pages = {1613-1630},
title = {Resonance Properties of Complex-Rotated Hamiltonians},
volume = {36},
journal = {Molecular Physics},
doi = {10.1080/00268977800102631}
}

@article{park2023feshbach,
  title={A Feshbach resonance in collisions between triplet ground-state molecules},
  author={Park, Juliana J and Lu, Yu-Kun and Jamison, Alan O and Tscherbul, Timur V and Ketterle, Wolfgang},
  journal={Nature},
  volume={614},
  number={7946},
  pages={54--58},
  year={2023},
  publisher={Nature Publishing Group UK London}
}

@article{higgott2019variational,
  title={Variational quantum computation of excited states},
  author={Higgott, Oscar and Wang, Daochen and Brierley, Stephen},
  journal={Quantum},
  volume={3},
  pages={156},
  year={2019},
  publisher={Verein zur F{\"o}rderung des Open Access Publizierens in den Quantenwissenschaften}
}

@article{Riss1993Calculation,
doi = {10.1088/0953-4075/26/23/021},
url = {https://dx.doi.org/10.1088/0953-4075/26/23/021},
year = {1993},
month = {dec},
publisher = {},
volume = {26},
number = {23},
pages = {4503},
author = {U V Riss and  H -D Meyer},
title = {Calculation of resonance energies and widths using the complex absorbing potential method},
journal = {Journal of Physics B},
}

@article{Muga2004Complex,
title = {Complex absorbing potentials},
journal = {Physics Reports},
volume = {395},
number = {6},
pages = {357-426},
year = {2004},
issn = {0370-1573},
doi = {https://doi.org/10.1016/j.physrep.2004.03.002},
url = {https://www.sciencedirect.com/science/article/pii/S0370157304001218},
author = {J.G. Muga and J.P. Palao and B. Navarro and I.L. Egusquiza},
}

@article{powell2009bobyqa,
  title={The {BOBYQA} algorithm for bound constrained optimization without derivatives},
  author={Powell, Michael JD},
  journal={Cambridge NA Report NA2009/06, University of Cambridge, Cambridge},
  volume={26},
  pages={26--46},
  year={2009}
}

@article{cartis2019improving,
  title={Improving the flexibility and robustness of model-based derivative-free optimization solvers},
  author={Cartis, Coralia and Fiala, Jan and Marteau, Benjamin and Roberts, Lindon},
  journal={ACM Transactions on Mathematical Software (TOMS)},
  volume={45},
  number={3},
  pages={1--41},
  year={2019},
  publisher={ACM New York, NY, USA}
}

@article{livny1997mechanisms,
  title={Mechanisms for high throughput computing},
  author={Livny, Miron and Basney, Jim and Raman, Rajesh and Tannenbaum, Todd},
  journal={SPEEDUP journal},
  volume={11},
  number={1},
  pages={36--40},
  year={1997}
}

@article{morgan2009high,
  title={High-throughput computing in the sciences},
  author={Morgan, Mark and Grimshaw, Andrew},
  journal={Methods in Enzymology},
  volume={467},
  pages={197--227},
  year={2009},
  publisher={Elsevier}
}

@article{thain2005distributed,
  title={Distributed computing in practice: the Condor experience},
  author={Thain, Douglas and Tannenbaum, Todd and Livny, Miron},
  journal={Concurrency and Computation: Practice and Experience},
  volume={17},
  number={2-4},
  pages={323--356},
  year={2005},
  publisher={Wiley Online Library}
}

@article{basney1999deploying,
  title={Deploying a high throughput computing cluster},
  author={Basney, Jim},
  journal={High Performance Cluster Computing},
  volume={1},
  year={1999}
}

@article{lorenz2025systematic,
  title={Systematic benchmarking of quantum computers: status and recommendations},
  author={Lorenz, Jeanette Miriam and Monz, Thomas and Eisert, Jens and Reitzner, Daniel and Schopfer, F{\'e}licien and Barbaresco, Fr{\'e}d{\'e}ric and Kurowski, Krzysztof and van der Schoot, Ward and Strohm, Thomas and Senellart, Jean and Perrault, C{\'e}cile M and Knufinke, Martin and Amodjee, Ziyad and Giardini, Mattia},
  journal={arXiv preprint arXiv:2503.04905},
  year={2025}
}

@article{moiseyev1998quantum,
  title={Quantum theory of resonances: calculating energies, widths and cross-sections by complex scaling},
  author={Moiseyev, Nimrod},
  journal={Physics Reports},
  volume={302},
  number={5-6},
  pages={212--293},
  year={1998},
  publisher={Elsevier}
}

@article{singh2025quantum,
  title={Quantum simulations of nuclear resonances with variational methods},
  author={Singh, Ashutosh and Siwach, Pooja and Arumugam, P},
  journal={Physical Review C},
  volume={112},
  number={2},
  pages={024323},
  year={2025},
  publisher={APS}
}

@article{cornish2024quantum,
  title={Quantum computation and quantum simulation with ultracold molecules},
  author={Cornish, Simon L and Tarbutt, Michael R and Hazzard, Kaden RA},
  journal={Nature Physics},
  volume={20},
  number={5},
  pages={730--740},
  year={2024},
  publisher={Nature Publishing Group UK London}
}

@article{son2022control,
  title={Control of reactive collisions by quantum interference},
  author={Son, Hyungmok and Park, Juliana J and Lu, Yu-Kun and Jamison, Alan O and Karman, Tijs and Ketterle, Wolfgang},
  journal={Science},
  volume={375},
  number={6584},
  pages={1006--1010},
  year={2022},
  publisher={American Association for the Advancement of Science}
}

@article{Tilly2022Variational,
title = {The Variational Quantum Eigensolver: A review of methods and best practices},
journal = {Physics Reports},
volume = {986},
pages = {1-128},
year = {2022},
issn = {0370-1573},
doi = {https://doi.org/10.1016/j.physrep.2022.08.003},
url = {https://www.sciencedirect.com/science/article/pii/S0370157322003118},
author = {Jules Tilly and Hongxiang Chen and Shuxiang Cao and Dario Picozzi and Kanav Setia and Ying Li and Edward Grant and Leonard Wossnig and Ivan Rungger and George H. Booth and Jonathan Tennyson}
}

@article{Nakanishi2020NFT,
   title={Sequential minimal optimization for quantum-classical hybrid algorithms},
   volume={2},
   ISSN={2643-1564},
   url={http://dx.doi.org/10.1103/PhysRevResearch.2.043158},
   DOI={10.1103/physrevresearch.2.043158},
   number={4},
   journal={Physical Review Research},
   publisher={American Physical Society (APS)},
   author={Nakanishi, Ken M. and Fujii, Keisuke and Todo, Synge},
   year={2020},
   month=oct }

@article{wang2010measurement,
  title={Measurement-based quantum phase estimation algorithm for finding eigenvalues of non-unitary matrices},
  author={Wang, Hefeng and Wu, Lian-Ao and Liu, Yu-xi and Nori, Franco},
  journal={Physical Review A},
  volume={82},
  number={6},
  pages={062303},
  year={2010},
  publisher={APS}
}

@article{daskin2014universal,
  title={A universal quantum circuit scheme for finding complex eigenvalues},
  author={Daskin, Anmer and Grama, Ananth and Kais, Sabre},
  journal={Quantum Information Processing},
  volume={13},
  pages={333--353},
  year={2014},
  publisher={Springer}
}

@article{bian2019quantum,
  title={Quantum computing methods for electronic states of the water molecule},
  author={Bian, Teng and Murphy, Daniel and Xia, Rongxin and Daskin, Ammar and Kais, Sabre},
  journal={Molecular Physics},
  volume={117},
  number={15-16},
  pages={2069--2082},
  year={2019},
  publisher={Taylor and Francis}
}

@article{parker2020quantum,
  title={Quantum phase estimation for a class of generalized eigenvalue problems},
  author={Parker, Jeffrey B and Joseph, Ilon},
  journal={Physical Review A},
  volume={102},
  number={2},
  pages={022422},
  year={2020},
  publisher={APS}
}

@article{shao2020quantum,
  title={Quantum algorithms for the polynomial eigenvalue problems},
  author={Shao, Changpeng and Liu, Jin-Peng},
  journal={arXiv preprint arXiv:2010.15027},
  year={2020}
}

@article{shao2022computing,
  title={Computing eigenvalues of diagonalizable matrices on a quantum computer},
  author={Shao, Changpeng},
  journal={ACM Transactions on Quantum Computing},
  volume={3},
  number={4},
  pages={1--20},
  year={2022},
  publisher={ACM New York, NY}
}

@article{zhao2023universal,
  title={A universal variational quantum eigensolver for non-Hermitian systems},
  author={Zhao, Huanfeng and Zhang, Peng and Wei, Tzu-Chieh},
  journal={Scientific Reports},
  volume={13},
  number={1},
  pages={22313},
  year={2023},
  publisher={Nature Publishing Group UK London}
}

@article{zhang2025quantum,
  title={Quantum computing for extracting nuclear resonances},
  author={Zhang, Hantao and Bai, Dong and Ren, Zhongzhou},
  journal={Physics Letters B},
  volume={860},
  pages={139187},
  year={2025},
  publisher={Elsevier}
}

@article{xie2024variational,
  title={Variational quantum algorithms for scanning the complex spectrum of non-Hermitian systems},
  author={Xie, Xu-Dan and Xue, Zheng-Yuan and Zhang, Dan-Bo},
  journal={Frontiers of Physics},
  volume={19},
  number={4},
  pages={41202},
  year={2024},
  publisher={Springer}
}

\color{black}
\appendix
\section{Appendix: Edge Cases of Exponential Zero-Noise Extrapolation\label{sec:EdgeCases}}

In edge cases for which exponential three-point ZNE produces a physically unrealizable zero-noise result $x_0\notin\mathcal{R}$, linear zero-noise extrapolation is employed to mitigate gate error as follows:

Where $x_{3}<x_{1}<x_{5}$, contrary to the physical expectation that
noise increases monotonically with $\lambda$, the optimal least-squares
curve with

\begin{equation}
A=\left(x_{1}+x_{3}\right)/2,\quad B\rightarrow0_{\pm},\quad C\rightarrow\infty\label{eq:x1Middle}
\end{equation}
remains nearly constant between $x_{1}$ and $x_{3}$ and increases
steeply to meet $x_{5}$ exactly; we thus exclude $x_{5}$ as an outlier
and employ the zero-noise estimate
\begin{equation}
x_{0}=\left(x_{1}+x_{3}\right)/2.
\end{equation}

Where $x_{1}<x_{5}<x_{3}$,\textbf{ }since the optimal least-squares
curve with
\begin{equation}
A=\left(x_{1}+x_{3}\right)/2,\quad B\rightarrow\pm\infty,\quad C\rightarrow-\infty\label{eq:x5Middle}
\end{equation}
yields the physically unreasonable estimate $x_0=\pm\infty$, and
since $x_{5}$ is expected to give rise to more error than $x_{3}$
where the system is not fully depolarized; we extrapolate $x_{1}$
and $x_{3}$ to $x_{0}$ via the linear fit $x_{\lambda}\approx A\lambda+B$
such that the zero-noise estimate is
\begin{equation}
x_{0}=\frac{3x_{1}-x_{3}}{2}.\label{eq:LinearFit}
\end{equation}

Where $x_{3}\approx x_{5}$, in order to avoid the instability $x_{0}\rightarrow\pm\infty$
as $x_{3}\rightarrow x_{5}$, we evaluate the statistical significance
between $x_{3}$ and $x_{5}$ according to a two-sample z-test at
the $\alpha=0.05$ level (for which statistical evidence corresponds
to a z-score of $|z|>1.96$). Given that measurement of the ancilla
projects the qubit onto either the $\left|0\right>$ or $\left|1\right>$
state, the $n$-shot standard deviation takes the form for a binomial
distribution
\begin{equation}
s=\sqrt{\frac{p(1-p)}{n}},
\end{equation}
which entails a joint standard deviation of
\[
s_{\text{joint}}=\sqrt{\frac{x_{3}(1-x_{3})+x_{5}(1-x_{5})}{n}}
\]
and associated z-score of
\[
z=\frac{x_{3}-x_{5}}{\sqrt{\frac{x_{3}(1-x_{3})+x_{5}(1-x_{5})}{n}}}.
\]
We therefore we use the exponential fit estimate Eq.~(\ref{eq:ExponentialFit})
if
\[
\frac{|x_{3}-x_{5}|}{\sqrt{\frac{x_{3}(1-x_{3})+x_{5}(1-x_{5})}{n}}}>1.96
\]
and the linear fit Eq.~(\ref{eq:LinearFit}) otherwise.

Where $x_{1}\approx x_{5}$, the estimate $x_{0}$ becomes undefined
(given the existence of two equally applicable but mutually inconsistent
least-squares curves Eq.~(\ref{eq:x1Middle}) and Eq.~(\ref{eq:x5Middle})),
such that we employ the average estimate for $x_{3}<x_{1}<x_{5}$
Eq.~(\ref{eq:x1Middle}) and $x_{1}<x_{5}<x_{3}$ Eq.~(\ref{eq:x5Middle})
\begin{equation}
x_{0}=x_{1}.
\end{equation}

Where $x_{1}\approx x_{3}\approx x_{5}$, the least squares curve
is
\begin{equation}
x_{1}=x_{3}=x_{5}=A+B,\quad C=0,
\end{equation}
such that we estimate $x_{0}$ as the constant value
\begin{equation}
x_{0}=x_{1}=x_{3}=x_{5}.
\end{equation}

\end{document}